\documentclass{aa}  

\usepackage{graphicx}
\usepackage{txfonts}
\begin{document}

   \title{The role of binaries in the enrichment of the early Galactic halo.}

   \subtitle{III. Carbon-enhanced metal-poor stars -- CEMP-$s$ stars}

   \author{T.T. Hansen
          \inst{1}
          \and
          J. Andersen\inst{2,3}\and B. Nordstr{\"o}m\inst{2,3}\and
          T. C. Beers\inst{4}\and V.M. Placco\inst{4}\and J. Yoon\inst{4}\and
          L.A. Buchhave\inst{5,6}} 

   \institute{Observatories of the Carnegie Institution of Washington, 813
     Santa Barbara St., Pasadena, CA 91101\\
              \email{thansen@carnegiescience.edu}
         \and
             Dark Cosmology Centre, The Niels Bohr Institute, University of Copenhagen,
             Juliane Maries Vej 30, DK-2100 Copenhagen, Denmark\\
             \email{ja@astro.ku.dk, birgitta@astro.ku.dk}
\and
    Stellar Astrophysics Centre, Department of Physics and Astronomy, Aarhus
    University, DK-8000 Aarhus C, Denmark 
\and
Department of Physics and JINA Center for the Evolution of the Elements,
University of Notre Dame, Notre Dame, IN 46556, USA\\ 
\email{tbeers@nd.edu,vplacco@nd.edu,jyoon4@nd.edu}            
\and
Harvard-Smithsonian Center for Astrophysics, Cambridge, MA 02138, USA
\and
Centre for Star and Planet Formation, University of Copenhagen,DK-1350 Copenhagen, Denmark \\
\email{buchhave@astro.ku.dk}
}

   \date{}

 
  \abstract
{Detailed spectroscopic studies of metal-poor halo stars have highlighted the
  important role of carbon-enhanced metal-poor (CEMP) stars in understanding
  the early production and ejection of carbon in the Galaxy and in
  identifying the progenitors of the CEMP stars among the first stars
  formed after the Big Bang. Recent work has also classified the CEMP stars by
  absolute carbon abundance, A(C), into high- and low-C bands, mostly
  populated by binary and single stars, respectively.}  
{Our aim is to determine the frequency and orbital parameters of binary systems 
  among the CEMP-$s$ stars, which exhibit strong enhancements of
  neutron-capture elements associated with the $s$-process. This allows us to
  test whether local mass transfer from a binary companion is necessary and
  sufficient to explain their dramatic carbon excesses.}  
{We have systematically monitored the radial velocities of a sample of 22 CEMP-$s$ 
  stars for several years with $\sim$monthly high-resolution, low S/N {\'e}chelle 
  spectra obtained at the Nordic Optical Telescope (NOT) at La Palma, Spain. 
From these spectra, radial velocities with an accuracy of
$\approx100$ m~s$^{-1}$ were determined by cross-correlation with optimized
templates.} 
{Eighteen of the 22 stars exhibit clear orbital motion, yielding a binary frequency 
  of 82$\pm$10\%, while four stars appear to be single (18$\pm$10\%). We thus
  confirm that the binary frequency of CEMP-$s$ stars is much higher than for normal metal-poor giants, but {\it not} 
  100\% as previously claimed. Secure orbits are determined for 11 of the
  binaries and provisional orbits for six long-period systems ($P > 3,000$
  days), and orbital circularisation time scales are discussed.} 
{The conventional scenario of local mass transfer from a former AGB binary 
companion does appear to account for the chemical composition of 
{\it most} CEMP-$s$ stars. However, the excess of C and $s$-process elements 
in {\it some single} CEMP-$s$ stars was apparently transferred to 
their natal clouds by an external
  (distant) source. This finding has important implications for our
  understanding of carbon enrichment in the early Galactic halo and some
  high-redshift DLA systems, and of the mass loss from extremely metal-poor AGB stars.} 


\keywords{Galaxy: formation -- Galaxy: halo -- Stars: chemically peculiar binaries: spectroscopic -- ISM: structure.}
   \maketitle
%

\section{Introduction}

\begin{table*}
\caption{Coordinates, photometry, and abundances for the CEMP-$s$ 
and CEMP-$r/s$ stars monitored for radial-velocity variation}
\label{tbl-1}
\centering
\begin{tabular}{lrrrrcrrrc}
\hline\hline
Stellar ID & RA (J2000) & Dec (J2000) & $V$ & $B-V$ & Ref$_{Phot}$ &$\mathrm{[Fe/H]}$ &
$\mathrm{[C/Fe]}$ & $\mathrm{[Ba/Fe]}$ & Ref$_{Abund}$\\
\hline
\multicolumn{10}{c}{CEMP-$s$}\\
\hline
\object{HE~0002$-$1037} & 00:05:23 &$-$10:20:23 & 13.70 & 0.48 &a&$-$3.75&$+$3.19&$+$1.67& 1 \\
\object{HE~0111$-$1346} & 01:13:47 &$-$13:30:50 & 12.48 & 1.31 &b&$-$1.91&$+$1.70&$<+$2.32&2,1\\
\object{HE~0151$-$0341} & 01:53:43 &$-$03:27:14 & 13.36 & 1.14 &b&$-$2.46&$+$2.46&$+$1.22 &2,1\\
\object{HE~0206$-$1916} & 02:09:20 &$-$19:01:55 & 14.00 & 1.13 &b&$-$2.09&$+$2.10&$+$1.97& 3\\
\object{HE~0319$-$0215} & 03:21:46 &$-$02:04:34 & 13.79 & 1.39 &b&$-$2.30&$+$2.00&$+$0.52 &1\\
\object{HE~0430$-$1609}$^*$ & 04:32:51 &$-$16:03:39 & 13.17 & 1.25 &b&$-$3.00&$+$1.14&$+$1.62 &1\\
\object{HE~0441$-$0652} & 04:43:30 &$-$06:46:54 & 14.23 & 1.02 &b&$-$2.47&$+$1.38&$+$1.11&3\\
\object{HE~0507$-$1430} & 05:09:17 &$-$16:50:05 & 14.49 & 1.54 &b&$-$2.40&$+$2.60&$+$1.30&4\\
\object{HE~0507$-$1653} & 05:10:08 &$-$14:26:32 & 12.51 & 1.13 &b&$-$1.38&$+$1.29&$+$1.89&3\\
\object{HE~0854$+$0151} & 08:57:30 &$+$01:39:50 & 14.98 & 0.92 &b&$-$1.80&$+$1.60&$+$0.82&1\\
\object{HE~0959$-$1424} & 10:02:04 &$-$14:39:22 & 13.37 & 0.60 &b&$-$1.42&$+$2.30&$+$1.24&1 \\
\object{HE~1031$-$0020} & 10:34:24 &$-$00:36:09 & 11.87 & 0.74 &a&$-$2.81&$+$1.58&$+$1.55&8\\
\object{HE~1045$+$0226} & 10:48:03 &$+$02:10:47 & 14.10 & 1.01 &a&$-$2.20&$+$0.97&$+$1.24&5\\
\object{HE~1046$-$1352} & 10:48:30 &$-$14:08:12 & 14.71 & 0.68 &b&$-$2.76&$+$3.30&$+$1.38&1\\
\object{CS~30301$-$015} & 15:08:57 &$+$02:30:19 & 13.04 & 1.00 &b&$-$2.64&$+$1.60&$+$1.45&7\\
\object{HE~1523$-$1155} & 15:26:41 &$-$12:05:43 & 13.23 & 1.35 &b&$-$2.15&$+$1.86&$+$1.72&3\\
\object{HE~2201$-$0345} & 22:03:58 &$-$03:30:54 & 14.31 & 1.18 &b&$-$2.80&$+$2.30&$+$0.62&1 \\
\object{HE~2312$-$0758} & 23:14:55 &$-$07:42:32 & 14.32 & 1.02 &a&$-$3.47&$+$1.86&$+$1.99&1 \\
\object{HE~2330$-$0555} & 23:32:55 &$-$05:38:50 & 14.56 & 0.85 &b&$-$2.78&$+$2.09&$+$1.22&3\\
\hline
\multicolumn{10}{c}{CEMP-$r/s$}\\
\hline
\object{HE~0017$+$0055} & 00:20:22 &$+$01:12:07 & 11.46 & 1.53 &b&$-$2.40&$+$2.17&$>+$1.99&10\\
\object{HE~0039$-$2635}$^{**}$ & 00:41:40 &$-$26:18:54 & 12.22 & 1.12 &b&$-$2.90&$+$2.63&$+$2.03&6\\
\object{LP~624$-$44}    & 16:43:14 &$-$01:55:30 & 11.68 & 1.16 &a&$-$2.72&$+$2.25&$+$2.83&9\\
\hline
\end{tabular}
\tablefoot{$^*$ = LP~775$-$30; $^{**}$ = CS~29497$-$034.}
\tablebib{{\it Photometry:} a) \citet{henden2015}, b) \citet{beers2007a}.   
{\it Abundances:} 1) This work, 2) \citet{kennedy2011}, 3) \citet{aoki2007}, 4)
\citet{beers2007b}, 5) \citet{cohen2013}, 6) \citet{barbuy2005}, 7)
\citet{aoki2002b}, 8) \citet{cohen2006}, 9) \citet{aoki2002c}, 10) \citet{jorissen2015a}.}
\end{table*}

Over the past few decades, large spectroscopic surveys have identified 
numerous very metal-poor (VMP; $\mathrm{[Fe/H]} < -2.0$) and extremely 
metal-poor (EMP; $\mathrm{[Fe/H]} < -3.0$) stars in the halo system of the 
Milky Way. High-resolution follow-up spectroscopy has also provided an
increasingly detailed picture of the star-to-star elemental-abundance
variations that constrain the early chemical evolution of the Galaxy
\citep[for reviews, see][]{beerschristlieb2005, ivezic2012,
frebelnorris2015}. The abundance patterns of individual chemically-peculiar
stars that deviate markedly from those of the bulk of Population II stars can
then be used to identify the nature of the progenitors and nucleosynthetic
processes responsible for the production of their distinctive chemical
signatures. Dilution of these signatures by later mixing with the interstellar
medium (ISM) of the Galaxy ultimately establishes the mean abundance trends
for relatively more metal-rich stars. 

A key element in this context is carbon, which is found to be
over-abundant in a large fraction of VMP stars ($\ga$20\% for
  $\mathrm{[Fe/H]}\leq -2$).
Carbon-enhanced metal-poor (CEMP) stars were originally identified among
the VMP and EMP stars discovered in the HK survey of Beers, Preston, \&
Shectman \citep{beers1985,beers1992} and the Hamburg/ESO survey of 
Christlieb and collaborators \citep{christlieb2008}, and supplemented by a 
number of surveys since. The fraction of CEMP stars rises with decreasing
 metallicity (conventionally tracked by the iron abundance,
$\mathrm{[Fe/H]}$); hence they are of particular importance for studies
of the early chemical evolution of the Galactic halo. 

The CEMP stars comprise a number of sub-classes
\citep[see][]{beerschristlieb2005}. The best-populated of these are the 
CEMP-$s$ and CEMP-no stars, characterised by the presence or absence of 
enhancements in $s$-process elements in addition to their carbon enhancement. 
The great majority of the former can be accounted for by scenarios 
involving transfer of enriched material from a binary companion that has 
passed through the asymptotic giant-branch (AGB) stage of evolution. 

The origin of the latter has still not been identified with certainty, but
their binary frequency is not higher than among metal-poor giants in general
 (see Paper~II of this series, \citealt{hansen2015c}). As discussed there,
a number of lines of evidence strongly suggest that the CEMP-no stars contain 
the nucleosynthesis products of the very first stars born in the Universe, 
i.e., that they are bona-fide second-generation stars. A third, less 
populated, sub-class is the CEMP-$r/s$ stars (which exhibit enhancements 
of both $r$-process and $s$-process elements in addition to that of carbon); 
their origin is presently poorly understood and needs further observational 
attention.

\citet{lucatello2005} carried out a limited multi-epoch RV-monitoring
survey of 19 CEMP-$s$ stars, and by combining with results from previous
authors, argued that some 68\% of the stars in their sample exhibited
evidence for radial-velocity variation. Based on the \citet{duquennoy91} 
distributions of orbital elements for Solar-type dwarfs and their simulation 
of the sampling of phase space by the velocity windows covered in their
observations, they concluded that the observations were compatible with 
100\% of CEMP-$s$ stars being members of binary (or multiple) systems.  
A re-analysis of this sample augmented with new data by \citet{starkenburg2014} 
came to a similar conclusion.  However, we emphasize that the size of
the sample considered, and the range of periods that could be examined
based on these data, is still relatively small.  

In this series of papers we present the results of an eight-year 
programme of precise radial-velocity monitoring and prompt, systematic
follow-up of potentially variable objects, for larger samples of
chemically-peculiar VMP and EMP stars than have heretofore received such close
attention. Our goal is to perform a solid test whether the distinctive
abundance signatures of these objects can be accounted for by alteration of
their birth chemistry by highly-evolved binary companions. 

\citet{hansen2011} first showed that the enhancement of $r$-process
elements observed in a small fraction (3-5\%) of VMP and EMP stars is {\it not
causally connected} to membership in a binary system, a conclusion that
was confirmed and further strengthened in Paper~I of this series
\citep{hansen2015b}. Paper~II \citep{hansen2015c} examined the same question 
for the class of CEMP-no stars and found  
that only 17$\pm$9\% (4 of 24) of their programme stars were binaries,
identical to the binary frequency found in
metal-poor red giants. The present Paper~III addresses the
extent to which binaries may play a role in the origin of CEMP-$s$ 
and CEMP-$r/s$ stars, using the same approach.

This paper is outlined as follows: Section 2 summarises the selection of
our programme stars and briefly describes our observational 
strategy and the techniques employed. Results are presented in Section 3, and
Section 4 describes the orbital properties of our binary programme stars. In
Section 5, we discuss the constraints imposed by these results on the
progenitors of CEMP-$s$ stars; a similar discussion for CEMP-$r/s$ stars
is provided in Section 6. Section 7 discusses the significance of the single 
stars identified in our programme, and Section 8 presents our conclusions 
and perspectives on what can be learned from future spectroscopic results on 
CEMP-$s$ and CEMP-$r/s$ stars.

\begin{table*}[th]
\caption{Barium and europium abundances for the potential CEMP-$r/s$ stars in
    the sample}
\label{tbl-Eu}
\centering
\begin{tabular}{lrrrl}
\hline\hline
Stellar ID & $\mathrm{[Ba/Fe]}$ & $\mathrm{[Eu/Fe]}$ & $\mathrm{[Ba/Eu]}$ & Ref\\
\hline
\object{HE~0017$+$0055}&>$+$1.9&$+$2.3   &>$-$0.4 & \citet{jorissen2015a}\\
\object{HE~0039$-$2635}&$+$2.03&$+$1.80   &$+$0.23 & \citet{barbuy2005}\\
\object{HE~1031$-$0020}&$+$1.55&$<+$0.82  &$>+$1.27& \citet{cohen2013}\\
\object{CS~30301$-$015}&$+$1.45&$+$0.20   &$+$1.25 & \citet{aoki2002b}\\
\object{LP~625$-$44}   &$+$2.83&$+$1.72   &$+$1.11 & \citet{aoki2002c}\\
\hline
\end{tabular}
\end{table*}


\section{Sample selection, observations, and analysis}

\subsection{Sample definition}

Our sample of stars is presented in Table \ref{tbl-1}, which lists their 
$V$ magnitudes, $B-V$ colours, and published $\mathrm{[Fe/H]}$, $\mathrm{[C/Fe]}$, 
and $\mathrm{[Ba/Fe]}$ abundances, either from the literature or determined as 
described below.

The majority of our programme stars are selected from the Hamburg/ESO
survey of Christlieb and collaborators \citep[HES;][]{christlieb2008}, with the
addition of one star from the HK survey, CS~30301$-$015. The CEMP star 
LP~625$-$44 was added since it is well-studied, and previous authors have 
suggested that it might be a CEMP-$r/s$ star. Two of the sample stars are
re-discoveries; HE~0039$-$2635 of the HK survey star CS~29497$-$034, and
HE~0430$-$1609 of the high proper-motion star LP~775$-$30. 

For the stars in Table \ref{tbl-1} labelled solely with a `1' in 
the final column ('this work'), the [Fe/H] and [C/Fe] abundances were determined from 
medium-resolution (R$\sim$2000) spectra, using the n-SSPP pipeline software 
\citep[described in detail by][]{beers2014}. These candidates were selected
from the CEMP candidate lists of \citet{placco2010,placco2011}, and earlier 
spectroscopic follow-up of HES candidates over the last 25 years. These stars 
would clearly benefit from higher-resolution spectroscopic abundance analyses. 

Additionally a number of stars in our programme had no abundance estimate (or upper
limit)  barium available in the literature; the $\mathrm{[Ba/Fe]}$
abundance is required in order to make a confident assignment of a star into
the CEMP-$s$ sub-class. For these stars we have derived Ba abundances (or upper
limits) from our co-added high-resolution spectra, following the procedure
described in Paper~II. This exercise clearly confirms the classification
of all of these stars as CEMP-$s$ stars (see Table \ref{tbl-1}).

Four stars in our sample, HE~0039$-$2635, HE~1031$-$0020, CS~30301$-$015,
and LP~625$-$44, have been suggested in the literature to be CEMP-$r/s$ stars:
Carbon stars showing enhancement in both $r$- and $s$-process elements
($0.0<\mathrm{[Ba/Eu]} < +0.5$; \citealt{beerschristlieb2005}). Additionally,
during the preparation of this paper, \citet{jorissen2015a} discovered that 
\object{HE~0017$+$0055} also has a very high Eu abundance. Table \ref{tbl-Eu}
lists the Ba and Eu abundances for all these stars, along with their
$\mathrm{[Ba/Eu]}$ ratios. The very high Eu abundance of HE~0017$+$0055,
combined with the lower limit on its Ba abundance, cause its $\mathrm{[Ba/Eu]}$
ratio to fall below the above formal limit. In summary, only HE~0039$-$2635
fully qualifies as a CEMP-$r/s$ star; nevertheless, Eu is detected in the
three other stars of the sample and should be accounted for in any formation
scenarios of the CEMP-$r/s$ stars, which are discussed more fully in
Sect. \ref{cemprs}.   

In the remainder of this paper, we retain the CEMP-$r/s$ label for
HE~0039$-$2635, HE~0017$+$0055 and LP~625$-$44. However, because
the Eu abundance of HE~1031$-$0020 is only an upper limit and
CS~30301$-$015 has only modestly-enhanced Eu, $\mathrm{[Eu/Fe]} =
+0.20$, we discuss these stars together with the other CEMP-$s$ stars.

\subsection{Observing strategy}

The key scientific goal of our project was to identify the single and 
binary stars in the sample and, if possible, determine the orbital periods 
and eccentricities with sufficient precision to understand the 
general properties of each class of stars. Accordingly, our observing 
strategy throughout the programme was to monitor the radial velocities 
of the sample stars regularly, precisely, and systematically in a homogeneous 
manner over a sufficiently long time span to detect any spectroscopic binaries 
among the stars, building on the examples of \citet{duquennoy91} and
\citet{carney2003}. 

Maintaining a roughly monthly cadence in the observations
was considered adequate for the expected long orbital periods. Aiming for a
precision of the individual observations of $\sim100$ m s$^{-1}$, and
continuing the observations for up to 2,900 days allowed us to detect orbital
motion with very long periods. Moreover, the observations were reduced and the
velocities inspected promptly after every observing night, so that any 
incipient variability could be detected and the observing cadence adapted 
as appropriate for each target. 

As described in Paper I, this strategy enabled us to identify the star 
\object{HE~1523$-$0901} as a very low-inclination binary despite its awkward 
period of 303 days and velocity semi-amplitude of only 0.35 km s$^{-1}$. This 
is evidence that we are able to securely detect binary orbits of even very
low amplitude, as seen also in the results displayed in Table
\ref{tbl-3}. However, observations of a given star were discontinued when
our key scientific objectives had been reached; spending precious telescope
time to achieve ultimate precision {\it per se} was not a priority beyond that
point. 

\subsection{Observations and data analysis}

Following the above strategy, the observations, reductions, and analysis 
procedures were the same as those of Papers~I and II of this series, to which 
the interested reader is referred for details; here we only give a short
summary. The stars were observed with the FIES spectrograph at the 2.5m Nordic
Optical Telescope (NOT). The spectra cover a wavelength range of
3640\,{\AA} to 7360\,{\AA}, at a resolving power of $R \approx46,000$ and
average signal-to-noise ratio (SNR) of $\approx$ 10. Background contamination
was minimised by observing the stars in grey time, when the cross-correlation
profile peaks of the stellar spectrum and any moonlight spectrum were
well-separated in velocity space. 

Reductions and multi-order cross-correlations were performed with software
developed by L. Buchhave. The template spectra employed for a given target 
were either the spectrum of the star with maximum signal (``Strongest''); 
a Co-added spectrum of all the best spectra (``Co-add''); a Synthetic
spectrum consisting of delta functions at the Solar wavelengths of the 
strongest stellar lines (``Delta''); or a co-added spectrum of a bright 
CEMP-$s$ star (HE~0507$-$1653) with a spectrum similar to that of the object.

Depending on the average quality of the spectra for a given star (S/N ratio; 
line density and strengths), the individual spectra were cross-correlated 
against one of these templates. The ``Strongest'' or ``Co-add'' templates 
were usually preferred, as they give a perfect match to the stellar spectrum and thus 
allow us to include the largest number of spectral orders in the correlations 
and optimise the precision of the derived radial velocities. However, for 
some low-signal spectra it was not possible to use these two templates; a 
Co-add spectrum of a bright CEMP-$s$ star was then used. 

The spectra of the CEMP-$s$ (and CEMP-$r/s$) stars are generally richer in 
strong lines than those of the $r$-process-enhanced and CEMP-no stars discussed 
in Papers~I and II. Therefore, most stars discussed here could be correlated with 
the ``Strongest'' or ``Co-add'' templates, the same template being used for 
all spectra of a given star. The typical accuracy of the resulting radial
velocities is 1$-$200 m s$^{-1}$.  

Finally, seven selected radial-velocity standard stars were monitored on 
every observing night throughout this programme. They are listed in Table~2 
of Paper~I, which gives the derived mean heliocentric velocities and standard 
deviations. The mean difference of our measured velocities for these stars 
from their standard values is 73~m~s$^{-1}$ with a standard deviation per 
star of 69~m~s$^{-1}$, demonstrating that our results are not limited by
the stability of the spectrograph.

\begin{table*}[ht]
\caption{Number of observations, adopted templates, mean heliocentric radial
  velocities and standard deviations, observed time spans, and
variability criterion $P(\chi^2)$ for the sample stars}
\label{tbl-2}
\centering
\begin{tabular}{lrlrrrcc}
\hline\hline
Stellar ID & N$_{obs}$ & Template & RV$_{mean}$ & $\sigma$  & $\Delta$T & 
$P(\chi^2)$ &Binary \\
           &      &          &(km-s$^{-1}$)&(km-s$^{-1}$)& (Days) \\
\hline
\multicolumn{8}{c}{CEMP-$s$}\\
\hline
\object{HE~0002$-$1037}&10&Co-add        & $-$31.295&5.957& 1066&0.000& Yes \\
\object{HE~0111$-$1346}& 9&Strongest     & $+$40.920&8.404& 1044&0.000& Yes \\ 
\object{HE~0151$-$0341}&11&Co-add        & $-$35.685&9.136& 1012&0.000& Yes \\ 
\object{HE~0206$-$1916}& 9&Co-add        &$-$199.536&0.121& 1044&0.233& No \\ 
\object{HE~0319$-$0215}&16&Co-add        &$-$225.782&2.357& 2207&0.000& Yes \\ 
\object{HE~0430$-$1609}&16&Co-add        &$+$231.821&1.727& 1184&0.000& Yes \\ 
\object{HE~0441$-$0652}&16&Co-add        & $-$30.647&2.655& 2371&0.000& Yes\\ 
\object{HE~0507$-$1430}&11&Strongest     & $+$44.802&7.920& 1064&0.000& Yes\\ 
\object{HE~0507$-$1653}&15&Co-add        &$+$348.280&4.859& 2124&0.000& Yes\\ 
\object{HE~0854$+$0151}&15&Co-add        &$+$138.297&7.798& 1757&0.000& Yes\\ 
\object{HE~0959$-$1424}&17&HE~0507$-$1653&$+$343.379&0.655& 2736&0.000& Yes\\ 
\object{HE~1031$-$0020}&22&Co-add        & $+$68.660&1.157& 2923&0.000& Yes\\
\object{HE~1045$+$0226}& 6&HE~0507$-$1653&$+$131.498&0.280&  803&0.223& No\\ 
\object{CS~30301$-$015}&18&Co-add        & $+$86.607&0.077& 2234&0.883& No\\
\object{HE~1046$-$1352}&12&Strongest     & $+$79.471&21.250&1812&0.000& Yes\\ 
\object{HE~1523$-$1155}& 9&Co-add        & $-$42.607&3.781&  502&0.000& Yes\\ 
\object{HE~2201$-$0345}&27&Co-add        & $-$55.927&3.525& 2943&0.000& Yes\\ 
\object{HE~2312$-$0758}&11&Co-add        & $+$32.981&3.176& 1066&0.000& Yes\\ 
\object{HE~2330$-$0555}&17&Co-add        &$-$235.124&0.231& 2573&0.543& No\\ 
\hline
\multicolumn{8}{c}{CEMP-$r/s$}\\
\hline
\object{HE~0017$+$0055}&28&Strongest     & $-$80.219&1.168& 2943&0.000& Yes \\ 
\object{HE~0039$-$2635}& 2&Strongest     & $-$47.739& 6.136&  278&0.000& Yes \\
\object{LP~625$-$44}   &28&Co-add        & $+$35.036& 3.348& 2667&0.000& Yes\\
\hline
\end{tabular}
\end{table*}

\section{Results}

\begin{figure*}[ht]
\centering
\includegraphics[scale=0.6]{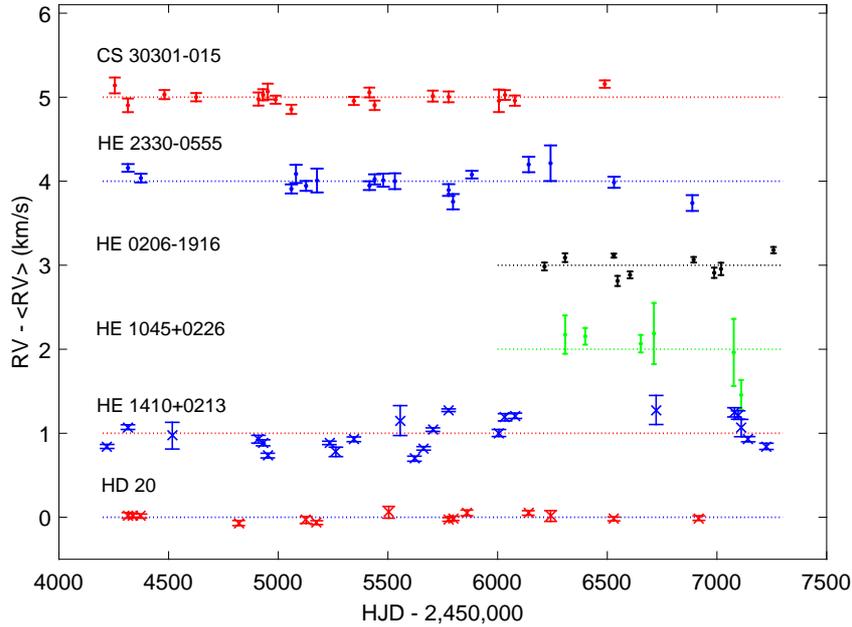}
\caption{Observed radial velocities for six constant stars from our
    project as functions of time, offset by 1 km-s$^{-1}$. 
    Top to bottom: CS~30301$-$015,
    HE~2330$-$0555, HE~0206$-$1916, and HE~1045$+$0226 (this paper);
    HE~1410$+$0213 (CEMP-no, Paper~II); and HD~20 ($r$-I, Paper~I).} 
\label{fig:constant}
\end{figure*}

The results of our radial-velocity monitoring of the sample of CEMP-$s$ and 
CEMP-$r/s$ stars in Table \ref{tbl-1} are summarised in Table \ref{tbl-2}, which 
lists the number of observations (N$_{obs}$) for each star, the cross-correlation
template used, the resulting mean heliocentric radial velocity (RV$_{mean}$) and
its standard deviation ($\sigma$), the observed time span ($\Delta$T), the
probability that the velocity is constant ($P(\chi^2)$), and our conclusion
whether the star is a binary. The individual radial-velocity observations and
their associated internal errors, computed as described in Paper I, are given
in Appendix~A. 

A few stars in our sample have radial velocities reported in the literature. 
The published measurements and total time spans covered for the four stars 
without significant radial-velocity variations are listed in Table \ref{tbl-B1}.
However, as these data are few in number and exhibit offsets of up to 
1 km-s$^{-1}$ between different sources that cannot be properly evaluated, we 
have not included them in our computations of $P(\chi^2)$ and the accompanying 
discussion. For the binary stars, the literature data are reviewed in Sect. 
\ref{binaries} and included in the orbital solutions when found useful. 

\subsection{Identifying the single and binary stars}

As in Paper~II, we have assessed the binary status of each individual 
star by calculating the standard $\chi^2$ parameter for variability to evaluate 
the probability, $P(\chi^2)$, that the radial velocity is constant within 
the observational errors. A 'floor error' or 'velocity jitter' of $\sim100$ 
m~s$^{-1}$ has been added in quadrature to the internal error to account for 
sources of external errors such as guiding or atmospheric dispersion 
and any intrinsic variability (discussed below). The resulting values of 
$P(\chi^2)$ are listed in Table \ref{tbl-2}, and demonstrate that at least 
eighteen of our programme stars exhibit highly significant radial-velocity 
variations over the eight-year period of monitoring, in most cases clearly 
due to orbital motion.
 
However, the context in which the computed $P(\chi^2)$ values are discussed
in this paper is the opposite of that in Papers~I and II. There, the underlying
hypothesis was that of a general population of constant-velocity (i.e., single)
stars with a uniform statistical distribution of $P(\chi^2)$ values between 0.01
and 1.00 (see \citealt{nordstrom1997}, Fig. 4, and \citealt{carney2003}, Fig. 2). 
Here, the default expectation, following \citet{lucatello2005} and
\citet{starkenburg2014}, is that {\it all} CEMP-$s$ (and CEMP-$r/s$) stars are
binaries (i.e., the frequency is 100\%).  

Thus, the simulations of \citet{lucatello2005} and \citet{starkenburg2014} 
did in fact only allow them to conclude that the compilation of data 
available to them (establishing an observed frequency of true binaries 
of 68$\pm$11\%) was {\it compatible with} that hypothesis, but it did not 
{\it prove} that it was {\it true}. Additional assumptions in their 
simulations were that {\it (i):} the distributions of
periods and other orbital parameters followed those derived by
\citet{duquennoy91} for Solar-type dwarfs, and {\it (ii):} all significant
radial-velocity variations were due to binary orbital motion.  

Our finding that four stars in our sample exhibit no signs of any binary 
orbital motion appears to contradict that assumption, especially since 18 of 
our 22 stars (82\%) are well-established binaries with $P(\chi^2)$ values 
below 10$^{-6}$, with secure or preliminary orbits for all but one -- an even 
{\it higher} fraction of actual binaries than found by \citet{lucatello2005}. 

Meanwhile, improvements in spectrograph design and radial-velocity
precision has revealed low-amplitude long-term velocity variations in
essentially all normal and especially bright giants. Accordingly, the
ability to detect binary (and exoplanet!) orbital motions of even very
low-amplitude and long- period binaries has not only improved due to
reduced observational errors, but it is also becoming increasingly limited by
intrinsic variability in the target star itself. Moreover, not all
potential binary companions are able to evolve past the AGB stage and
transfer $s$-process enriched material to the surviving star, so only
binary systems with companions of present-day mass in the white dwarf
range of 0.5-1.4~M$_{\sun}$ \citep{merle2015} need concern us here. We
therefore discuss the binary status of our potentially single stars
separately in the following section.

\subsection{The apparently single stars}
\label{singles}

The four stars in Table \ref{tbl-2} that are not established
binaries all have $P(\chi^2)$ values well above the limit of $P(\chi^2)$
= 0.05, beyond which \citet{carney2003} considered the stars to be safely
single and demonstrated that the distribution of their $P(\chi^2)$ values
is flat, as expected \citep[see also][, Fig. 4]{nordstrom1997}. However, as also 
shown by \citet{lucatello2005}, a value of $Q(\chi^2)$ (= 1 $-$ $P(\chi^2)$)  
$\ga$0.02 actually indicates that most of them are in fact more likely 
to be variable than constant. This by itself does not prove that any such 
variation is necessarily caused by binary {\it orbital motion}; it must also 
exhibit a characteristic, significant, systematic, and well-sampled pattern 
(see, e.g., \citealt{morbeygriffin1987}) . 

This is illustrated in Figure \ref{fig:constant}, which compares the
time histories of the velocities of the four single CEMP-$s$ stars in
our sample with those of the single stars HE~1410$+$0213 (a CEMP-no star,
Paper~II) and HD~20 (an $r$-I star, Paper~I). Three of these stars
(CS~30301$-$015, HE~0206$-$1916, and HD~20) have standard deviations of
$\la$100 m~s$^{-1}$ over their total periods of observation, and none of
them exhibits any sign of orbital motion.  

As discussed in detail in Paper~II, \object{HE~1410$+$0213} was initially
suspected of showing orbital motion with a period near 341 days and semi-amplitude 
$\la$ 300 m~s$^{-1}$, but it did not continue this behaviour and eventually was 
judged to be a single, pulsating star. From the sample presented in this paper, 
\object{HE~0017$+$0055} is found to be a long-period binary (see Table 
\ref{tbl-2}), but it also exhibits an additional regular velocity variation
of period $\approx$ 385 days, $e\approx$ 0.15, and semi-amplitude $K\approx$ 540 
m~s$^{-1}$. Finally, in Paper~I we found the highly $r$-process enhanced EMP star
\object{HE~1523$-$0901} to be a spectroscopic binary with a period of 303 days
and semi-amplitude 350 m~s$^{-1}$.   

Adopting these periods and amplitudes, modest orbital eccentricities,
and setting $M_1\approx0.8$ M$_{\sun}$ and $M_2\approx0.5-1.4$ 
M$_{\sun}$ for the observed star and the presumed white
dwarf companion, respectively, leads to orbital inclinations in the
range $1.5\degr-2\degr$. Figure \ref{fig:constant} suggests that any
undiscovered binaries among our four 'single' stars would have periods
in the range 1,000$-$10,000 days or even longer and semi-amplitudes
$\la100$ m~s$^{-1}$, leading to similarly low orbital inclinations.
Assuming a random distribution of orbits in space, the probability of
finding even one such closely face-on orbit is $\approx10^{-4}$, or less than 
1\% for our total sample of 63 stars. Having found HE~1523$-$0901 must then 
already be considered lucky; finding four such cases strains credulity. 

Continued radial-velocity monitoring might still reveal orbital motion in one 
of our 'constant' stars, notably in HE~1045$+$0226, but for now, we retain four 
as the most likely number of single stars, a fraction of $18\pm10$\%. Thus, the 
great majority of the CEMP-$s$ stars are still in binaries, but exceptions to 
the local mass-transfer scenario for their origin do appear to exist, as is the 
case for the class of CEMP-no stars discussed in Paper~II. Alternative scenarios 
are discussed in Sect. \ref{singlestars}. 

It is remarkable that two of our CEMP stars (HE~1410$+$0213 and
HE~0017$+$0055) appear to exhibit near-periodic low-amplitude velocity
variations of periods similar to those identified by \citet{riebel2010}
in the large OGLE data set of pulsating LMC giants and C-rich AGB stars.
Signatures of similar pulsations in another two CEMP stars are found in
the sample of \citet{jorissen2015b}, but not among carbon-normal VMP/EMP
$r$-process-enhanced stars. This suggests that the high molecular
opacities of C-rich stellar atmospheres may be their source, but photometric
confirmation of their existence and probable long periods in field stars is
difficult from the ground. 

Compounding this difficulty is the lack of reliable distances and absolute 
luminosities for isolated CEMP stars, and spectroscopic log $g$ values are 
generally uncertain guides due to inadequate resolution of these line-packed 
spectra. However, the impending precise trigonometric parallaxes and parallel 
uniform, precise, and well-sampled photometry from the Gaia mission should 
shortly put our understanding of the properties and evolution of these stars 
on a much safer footing.

\begin{table*}[th] 
\caption{Orbital parameters for the binary systems in our sample (mean
errors are given below each parameter}
\label{tbl-3}
\centering
\resizebox{\hsize}{!}{
\begin{tabular}{lrrrrrrrrrrr}
\hline\hline
Parameter & Period & $T_0$ & $K$ & $\gamma$ & $e$ & $\omega$ & $a$ sin$i$ 
          & $f(m)$ & $M_2$ & $R_{Roche}$ & $\sigma$ \\
Units     & (days) & (HJD) & (km~s$^{-1}$) & (km~s$^{-1}$) & & ($\degr$) & 
     (R$_{\sun}$) & (M$_{\sun}$) & (M$_{\sun}$) & (R$_{\sun}$) & (km~s$^{-1}$)\\
\hline
\multicolumn{12}{c}{CEMP-$s$ Stars}\\
\hline
\object{HE~1046$-$1352}&20.156 &51,199.88 & 30.19& $+$75.37& 0.00& 0 & 12.03& 0.057& 0.4& 5.8& 0.82 \\
      & 0.001& 0.04& 0.30& 0.16& 0.00& --- & 0.01& 0.008& 1.4& 22.0 \\
\object{HE~1523$-$1155}&309.34& 57,009.3& 5.21& $-$42.94& 0.272& 100& 30.66 & 0.0040& 0.4& 31.9& 0.018\\
      & 0.62& 2.6& 0.04& 0.11& 0.008& 2& 0.18& 0.0002& 1.4& 132.5\\
\object{HE~0151$-$0341}&359.07& 55,313.4& 12.15& $-$37.74& 0.00& 0  & 86.22& 0.0667& 0.45& 42.2 & 0.08 \\
      & 0.21& 0.7& 0.03& 0.03& 0.00& --- & 0.16& 0.0004& 1.4& 146 \\
\object{HE~0854$+$0151}& 389.85&55,305.58& 12.86& $+$133.58& 0.00& 0& 99.1& 0.0859& 0.5& 52.0& 0.13 \\
      & 0.07& 0.07& 0.03& 0.02& 0.00 & --- & 0.1& 0.0003& 1.5& 155 \\
\object{HE~0111$-$1346}& 403.81& 56,320.2& 12.74& $+$37.75& 0.00$^*$& 0 & 101.32& 0.8855& 0.55& 56.7 & 0.11 \\ 
      & 0.14& 0.1& 0.02 & 0.02 & --- & --- & 0.08&0.0001&1.4&162 \\
\object{HE~0507$-$1653}& 404.18& 55,840.24 & 7.090& $+$349.843& 0.00  & 0& 56.63& 0.0145& 0.4& 36& 0.20 \\
      & 0.05& 0.07& 0.008 & 0.006& 0.00 & --- & 0.03& 0.0002& 1.4& 156\\ 
\object{HE~0507$-$1430}& 446.96& 55,272.71& 10.927 & $+$42.961 & 0.0058  & 84 & 96.64 & 0.0604& 0.44& 47& 0.05 \\
      & 0.15 & 0.16 & 0.011 & 0.009 & 0.0015 &11& 0.06&0.0001&1.4& 169\\ 
\object{HE~0002$-$1037}& 740.9& 56,622.5& 9.50& $-$32.49& 0.142& 85& 138.0 & 0.064& 0.4& 67& 0.29 \\
      & 1.0& 0.5& 0.08& 0.03 & 0.005& 3& 0.5& 0.003& 1.4& 242 \\
\object{HE~2312$-$0758}& 1,890& 56,536& 4.33& $+$33.16& 0.26$^*$& 0& 156& 0.014& 0.4& 99& 0.14 \\
      & 57& 7& 0.08& 0.15& --- & --- & 6& 0.005& 1.4& 447 \\ 
\object{HE~0319$-$0215}& 3,078& 53,572& 4.28& 227.23& 0.00& 0 & 260.3& 0.025& 0.4& 142& 0.34 \\
      & 25& 19& 0.05& 0.05& 0.00& --- & 3.2& 0.05& 1.4& 634 \\
\object{HE~1031$-$0020}& 3,867& 56,006& 1.78& $+$68.22& 0.38$^*$& 245 & 126& 0.002& 0.6& 284& 0.17 \\
      & 175& 54& 0.07& 0.06& ---& ---& 28& 0.035 \\
\object{HE~0430$-$1609}& 4,368& 58,873& 3.80& $+$231.15& 0.0$^*$& 0& 328& 0.025 & 0.6& 310& 0.04 \\
      & 198 &96& 0.15& 0.06& ---& ---& 20& 0.002 \\
\object{HE~0441$-$0652}& 5,223& 58,408& 8.7& $-$34.58& 0.48$^*$& 217  & 785& 0.24& 0.72& 540& 0.64 \\
      & 628& 308& 3.3& 1.95& ---& ---& 193& 0.26 \\
\object{HE~2201$-$0345}& 10,093& 55,696 & 5.13& $-$58.08& 0.67& 28  & 749& 0.055& 0.6& 556& 0.17 \\ 
      & 1,656& 8& 0.05& 0.24& 0.04& 1& 129& 0.029 \\ 
\hline
\multicolumn{12}{c}{CEMP-$r/s$ Stars}\\
\hline
\object{HE~0039$-$2635}& 3,223& 51,900& 6.9 & $-$46.5 & 0.46$^*$ & 239 & 424 & 0.09& 0.6 & 264& 1.32 \\
      & 36 & 12 & 0.5 & 10.4 & --- & --- & 139& 0.53 \\ 
\object{HE~0017$+$0055}& 3,529& 55,407& 1.57& $-$80.39& 0.43& 312 & 99& 0.001& 0.6& 64& 0.33 \\
      &236& 40& 0.06& 0.05& 0.05& 8& 9& 0.035 \\
\object{LP~625$-$44}& 4,863& 56,007& 6.35& $+$33.63& 0.35$^*$& 245     & 571& 0.10& 0.6& 332& 0.42 \\
      & 12& 16& 0.04& 0.06& --- & --- & 18& 0.01 \\
\hline
\end{tabular}}
\tablefoot{$^*$ Eccentricity fixed in the solution.}
\end{table*}

\subsection{Binary stars}
\label{binaries}
Radial velocities have been published for a number of our stars that have 
been found to exhibit variable radial velocities. These data have been included 
in our discussion of certain or potential binaries in the following. \\

\noindent {\it CEMP-s stars:}

\begin{itemize}

\item{{\it \object{HE~0111$-$1346}} and {\it \object{HE~0507$-$1653}} were 
also observed extensively by \citet{jorissen2015b}, with results agreeing 
with ours within the errors.}\\ 

\item{{\it \object{HE~0430$-$1609}} and {\it \object{HE~1523$-$1155}} have 
suspiciously low rms errors for their orbital solutions. For HE~0430$-$1609, 
we can fit eccentric 2,000-day and circular 4000-day orbits to our data, with 
identical rms errors of only 40 m s$^{-1}$(!), but consider the latter more 
reliable (see end note). For HE~1523$-$1155, the velocities we derived from
our observations from 2014 and 2015 plus the observation by \citet{aoki2007} 
constrain the period accurately; in particular, much longer periods are excluded 
despite the significant orbital eccentricity of $e=0.3$. More
observations of these two stars are needed in order to determine the final
orbits with confidence.}\\ 

\item{{\it \object{HE~0507$-$1653}} and {\it \object{HE~1523$-$1155}}:
  \citet{aoki2007} has reported single radial velocities for these two objects 
 (353.0 km~s$^{-1}$ and $-$45.0 km~s$^{-1}$, respectively), which fit the
 orbital solutions derived from our own data; see Figure \ref{fig:orbit1}.} \\

\begin{figure}[ht]
\includegraphics[scale=0.6]{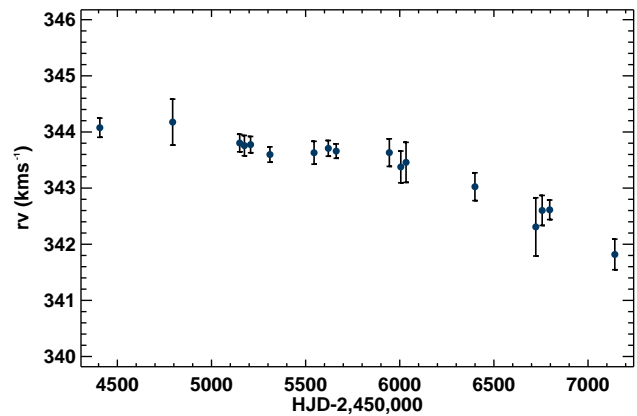}
\caption{Radial velocities measured for HE~0959$-$1424 as function of time,  
showing a clear decrease over the observing period.}
\label{fig:HE0959}
\end{figure}

\item{{\it \object{HE~0959$-$1424}} undoubtedly has a variable velocity, as 
indicated by the values of $\sigma$ and $P(\chi^2)$ given in Table \ref{tbl-2}.
Figure~\ref{fig:HE0959} shows the measured radial velocities as a
function of time. The slow, systematic velocity decrease 
suggests that \object{HE~0959$-$1424} is a binary with a very long
period, possibly of the order of 10,000 days or even (much) longer, and 
a semi-amplitude of a few km~s$^{-1}$. Low-amplitude short-term velocity
oscillations of the type seen in \object{HE~0017+0055} are not observed in
this star.} 

\end{itemize}

\noindent {\it CEMP-r/s stars:}

\begin{itemize}

\item{{\it \object{HE~0017$+$0055}} was monitored extensively in parallel with 
this programme by \citet{jorissen2015a}, who also discovered its large 
enhancement of Eu and other $r$-process elements. We have therefore moved it 
to the CEMP-$r/s$ subgroup of the sample. Moreover, the analysis of our
joint data set revealed a low-amplitude, short-period oscillation
superimposed on a highly significant long-term trend. For the short-term
oscillation, a Keplerian orbit was formally derived from data covering
$\approx$ 8 full cycles, but the tiny $f(m)$ of (6$\pm$1)
$\times$10$^{-6}$ M$_{\sun}$ implied an implausibly low orbital
inclination of $\approx2\degr$ (see Sect. \ref{singles}). 

An alternative interpretation in terms of stellar pulsations, based on the 
complete material, was therefore discussed by \citet{jorissen2015a}, who 
found the oscillations in this star to be similar to those in the 
CEMP-no star HE~1410$+$0213 (Paper~II). In the end, regardless of
the ultimate cause of these short-period, low-level oscillations, the
reality of the long-term orbital motion of HE~0017$+$0055 is not in
doubt, and we retain the star as a binary in our sample.} \\

\item{{\it \object{HE~0039$-$2635}} (alias {\it\object{CS~29497$-$034}}) was
  observed by both \citet{lucatello2005} and \citet{barbuy2005}. The former
  report one measurement from 2002, while the latter reported 11 independent
  measurements over a span of $\approx$ 3,000 days between 1995 and 2004 with
  an rms of 3.4 km~s$^{-1}$, and derived an orbital solution with $P = 4,130$
  days and $e = 0.2$. We have combined our results with the published data and
  find a more eccentric orbit with a shorter period ($P = 3,223$ days, $e
  =0.46$); see Figure \ref{fig:orbit2} and Table \ref{tbl-3}.}\\

\item{{\it \object{LP~625$-$44}} has published radial-velocity data from 
the following sources: \cite{norris1997}: Five observations during  
1988$-$1996; \citet{aoki2000}: Two observations from 1998$-$2000; and
\citet{lucatello2005}: Three observations from 2000$-$2002. None of
these authors had sufficient data to compute an orbital solution for
this star. However, combining their data with our own extensive series of 
measurements, and applying offsets between the literature velocities and
ours (Norris et al., $-$0.14 m~s$^{-1}$; Aoki et al., $+$451.96
m~s$^{-1}$; and Lucatello et al., $+$155.87 m~s$^{-1}$), we could
construct a data set covering a total time span of 9,582 days. From
this, we have computed an orbit with a period of $P = 4,863$ days
and $e$ = 0.35; see Table \ref{tbl-3} and Figure~\ref{fig:orbit2}.}  

\end{itemize}

\section{Binary orbital solutions}

\begin{figure*}
\centering
\includegraphics[scale=0.7]{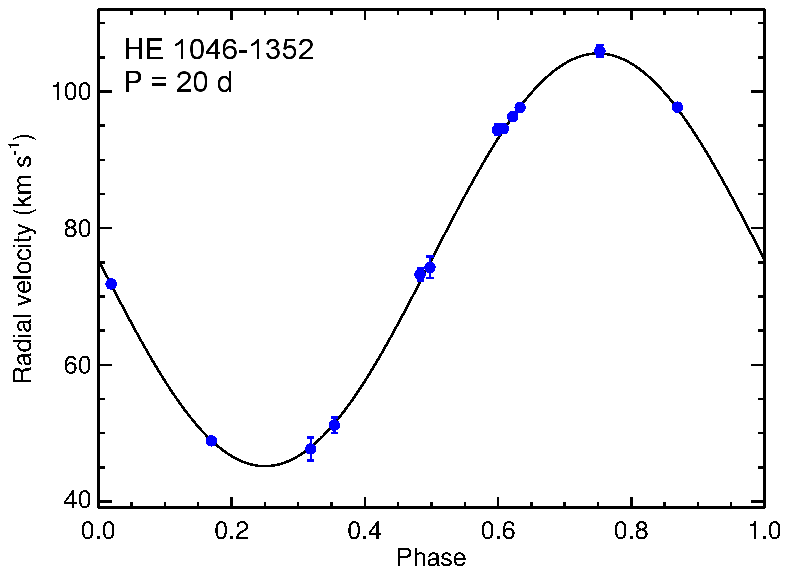}
\includegraphics[scale=0.7]{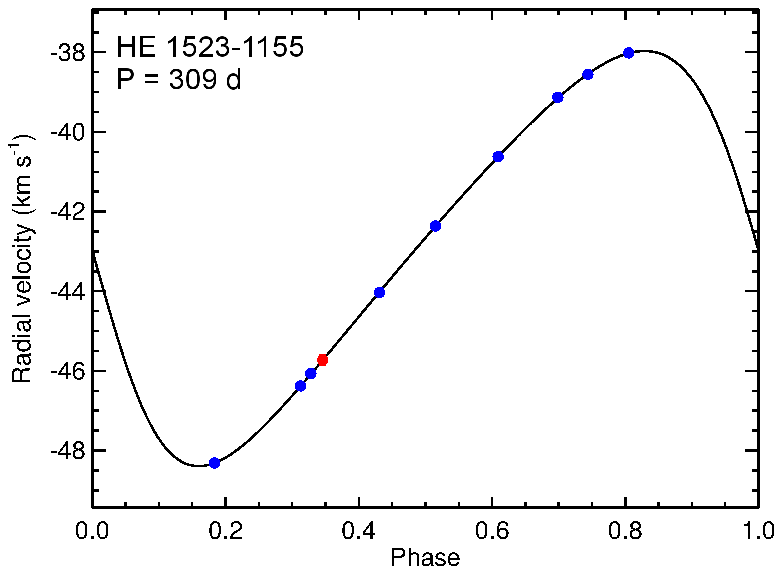}
\includegraphics[scale=0.7]{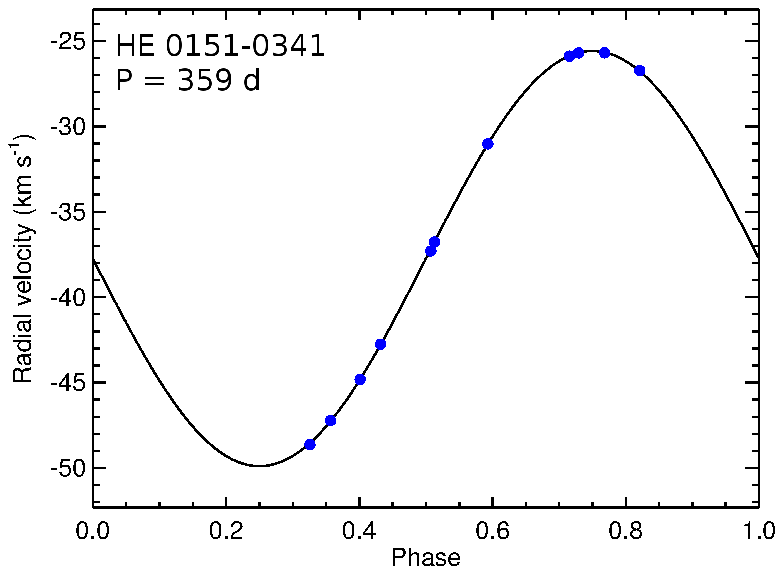}
\includegraphics[scale=0.7]{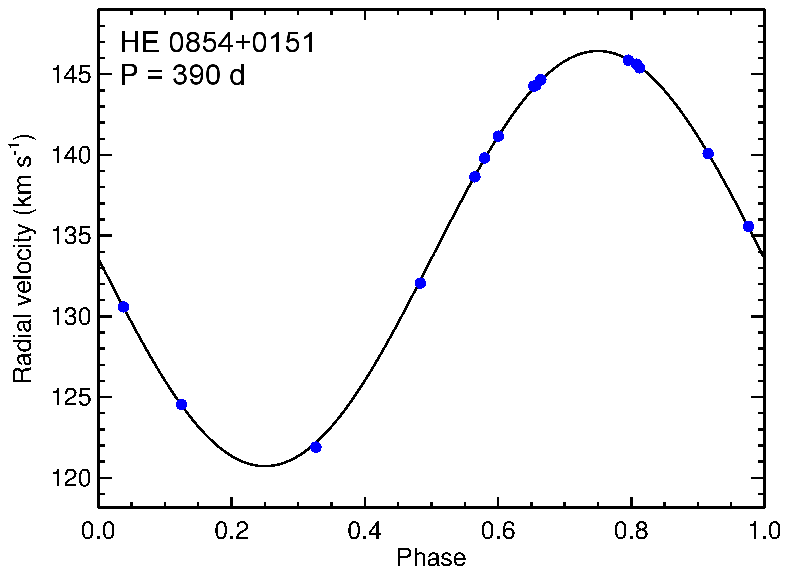}
\includegraphics[scale=0.7]{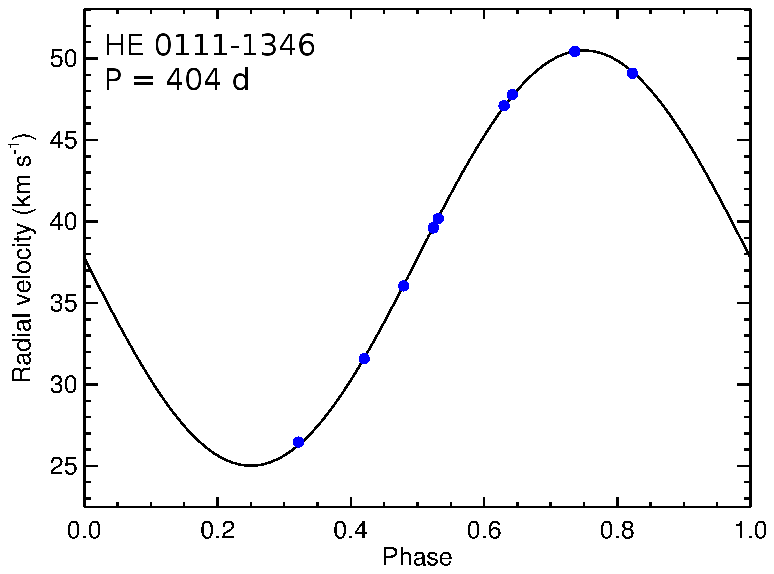}
\includegraphics[scale=0.7]{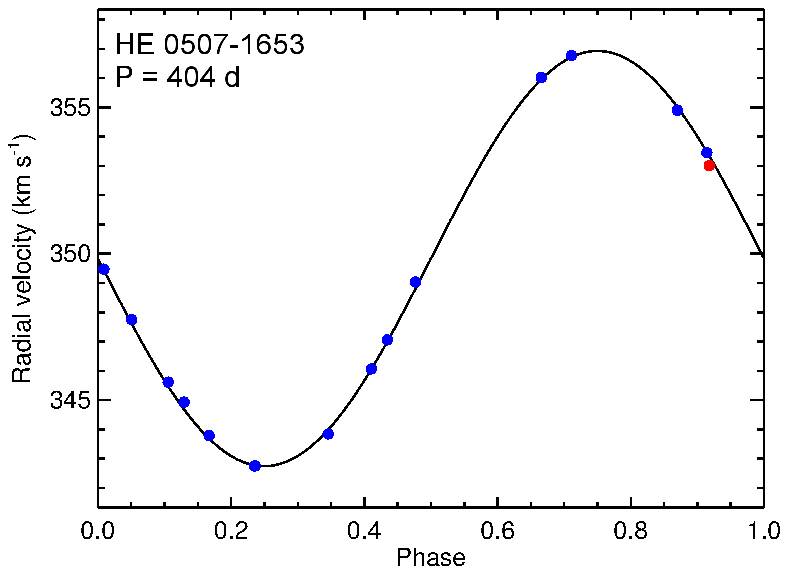}
\includegraphics[scale=0.7]{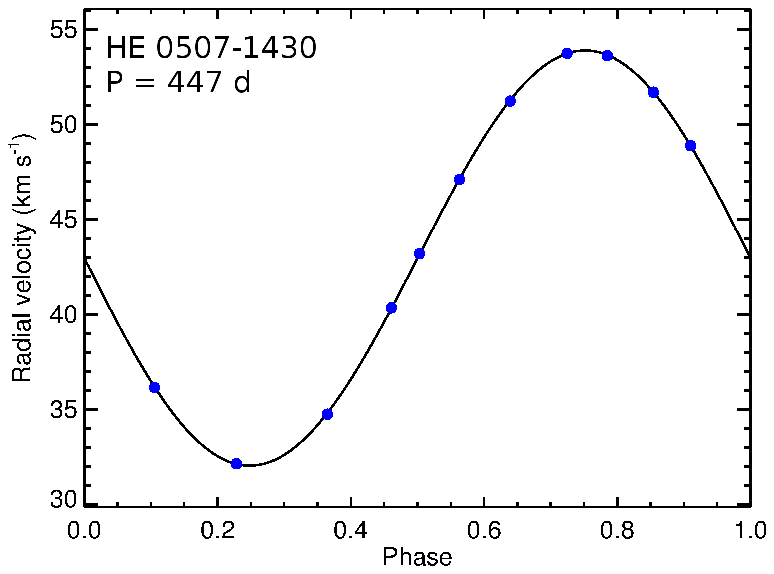}
\includegraphics[scale=0.7]{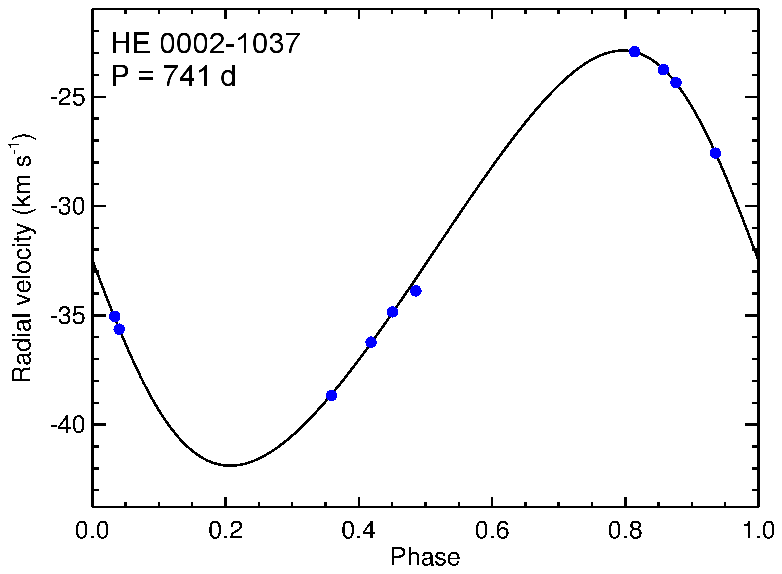}
\includegraphics[scale=0.7]{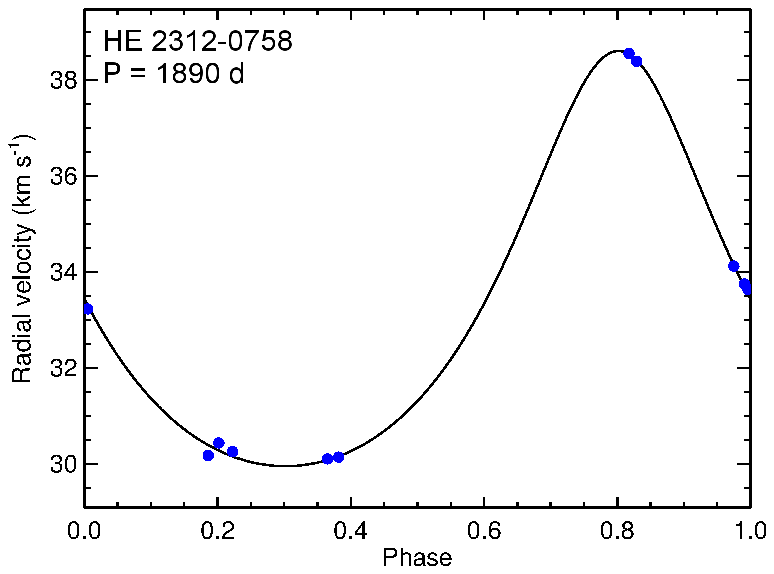}
\includegraphics[scale=0.7]{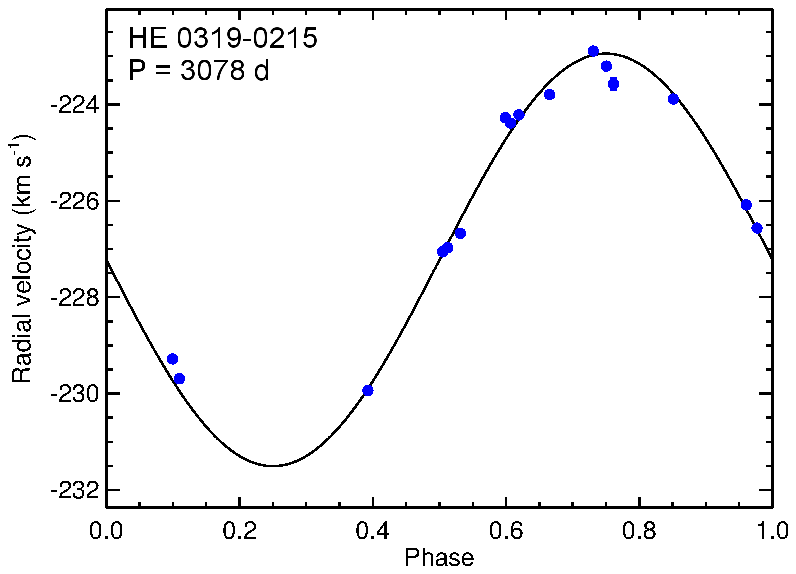}
\includegraphics[scale=0.7]{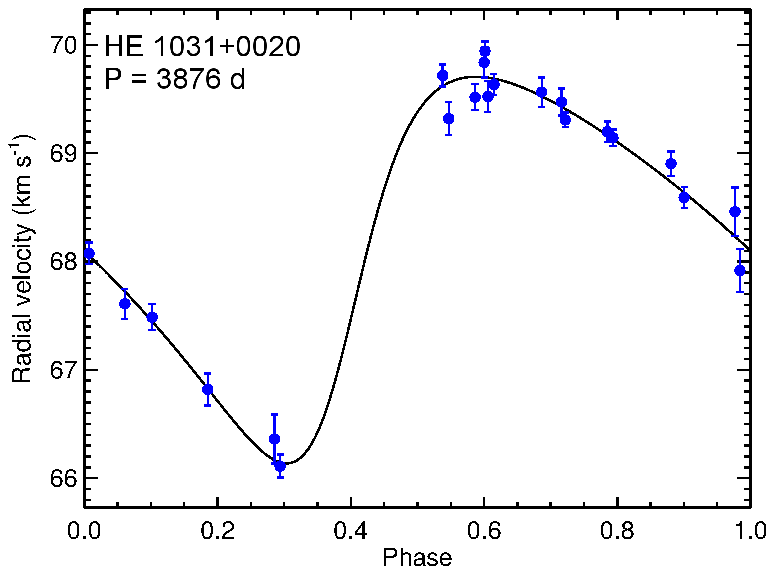}
\includegraphics[scale=0.7]{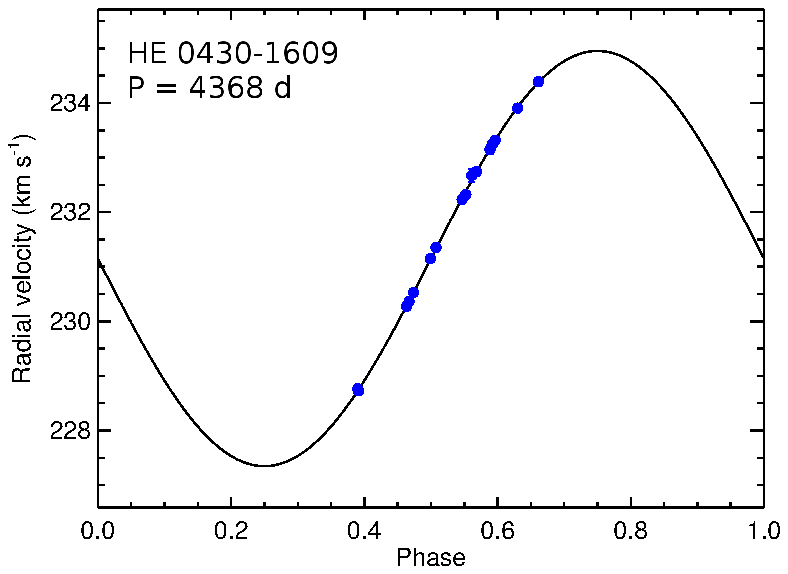}
\includegraphics[scale=0.7]{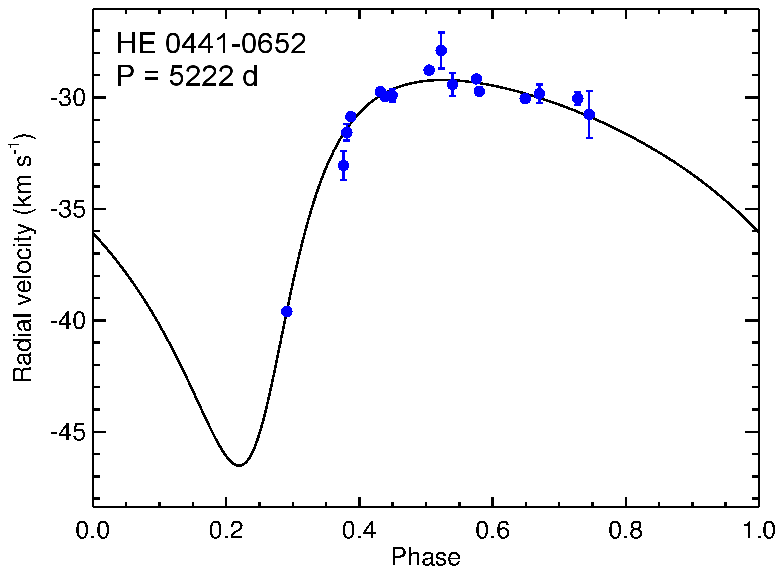}
\includegraphics[scale=0.7]{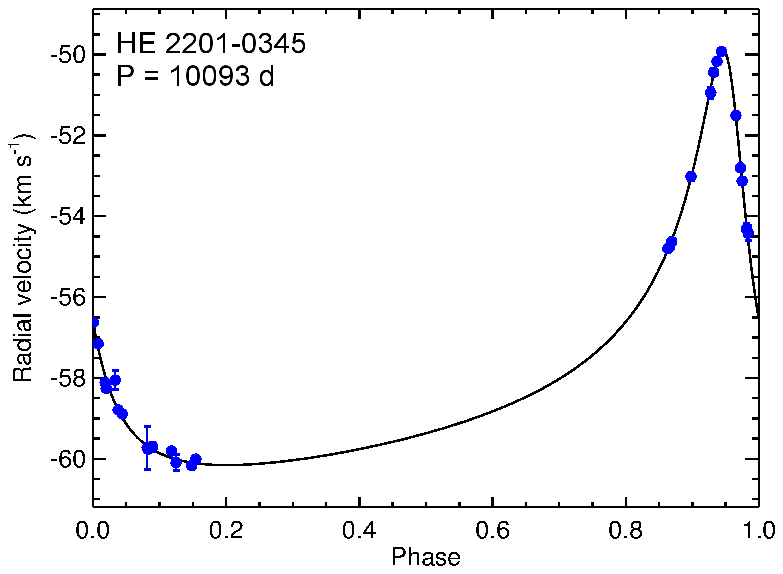}
\caption{Orbit solutions for the CEMP-$s$ binaries in our programme.  Blue
symbols: this work; red symbols: \citet{aoki2007}.}  
\label{fig:orbit1}
\end{figure*}

\begin{figure}
\centering
\includegraphics[scale=1.05]{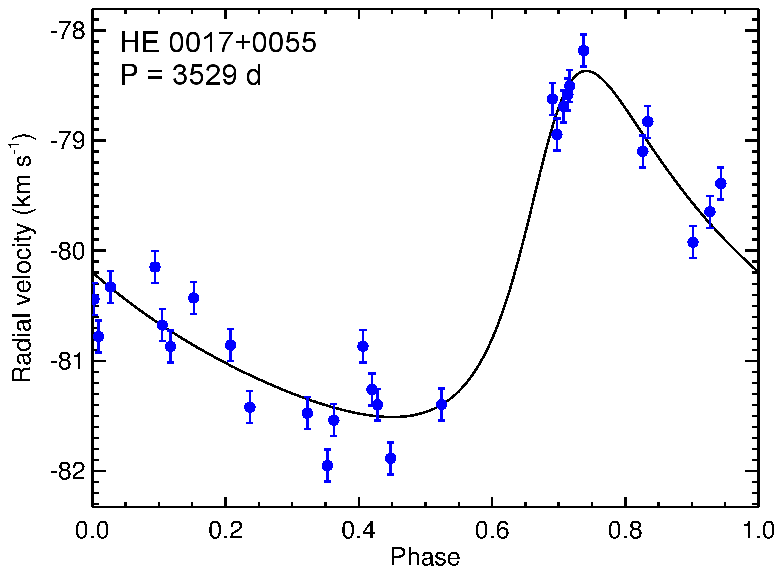}
\includegraphics[scale=0.15]{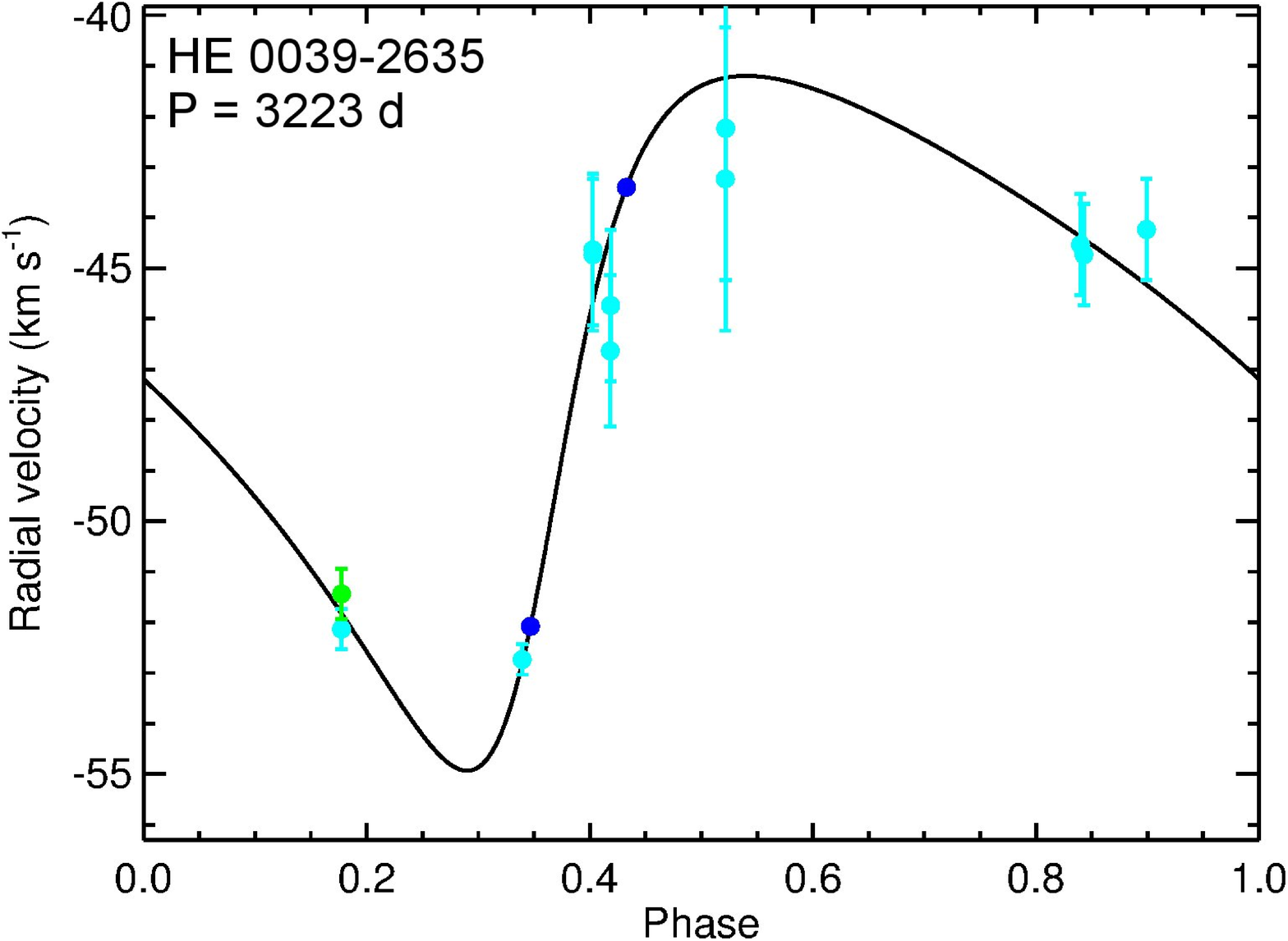}
\includegraphics[scale=0.15]{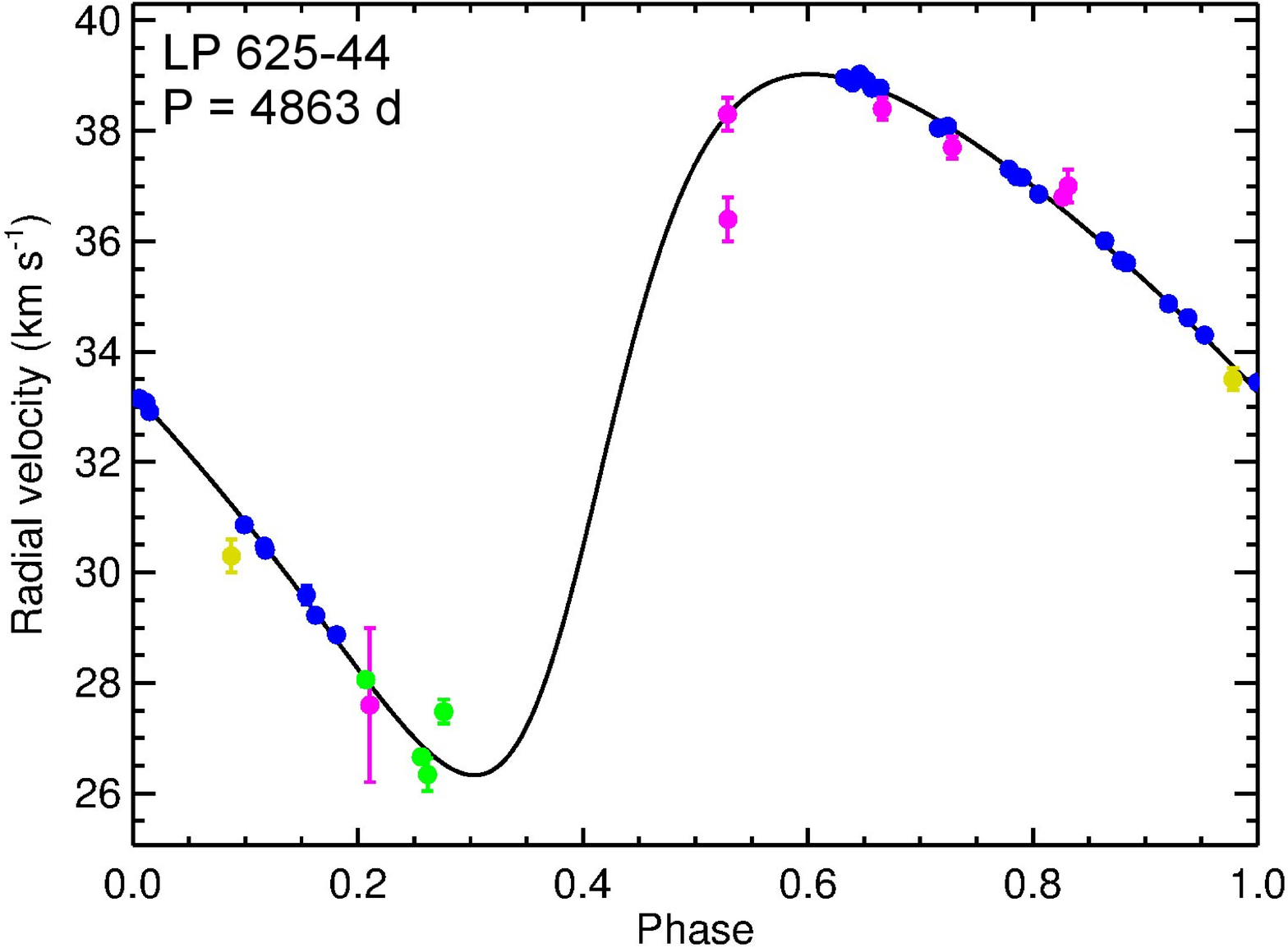}
\caption{Orbits for the CEMP-$r/s$ stars. Symbols: Blue: this work;
magenta: \citet{norris1997}; green: \citet{lucatello2005};
yellow: \citet{aoki2000}; light blue: \citet{barbuy2005}.}  
\label{fig:orbit2}
\end{figure}

Orbital solutions have been obtained for the 17 confirmed binary systems
in our sample, except the probable (very) long-period binary
HE~0959$-$1424, discussed above. Our final orbital parameters for
these 17 systems are listed in Table \ref{tbl-3} in order of increasing
period, and radial-velocity curves are shown, also in order of
increasing period, in Figure~\ref{fig:orbit1} for the CEMP-$s$ stars and
in Figure~\ref{fig:orbit2} for the CEMP-$r/s$ stars.

\clearpage 

The orbital periods of the binaries among our programme stars range from 20
days to $\approx$ 30 years, the majority having periods around one year; most
of these are in circular orbits. Naturally, for orbits with periods in excess
of the $\approx$ 3,000-day time span of our observations, the orbital
parameters have correspondingly larger errors.

Table \ref{tbl-3} also lists the Roche-lobe radii of the secondary stars
in these systems, calculated by the procedure described in Paper~I. For this, 
we have assumed a mass of 0.8~M$_\odot$ for the observed metal-poor giant
primary stars, and secondary masses of 0.4~M$_\odot$ (larger if
$i~=~90\degr$ is reached) and 1.4~M$_\odot$, respectively (minimum and 
maximum masses for the presumed white dwarf companion). For orbital periods
$P \ga$ 3,000 days, these Roche-lobe radii become very uncertain; therefore,
only indicative values for $M_2$~=~0.6~M$_{\sun}$ are given.

\subsection{Binary fraction and distribution of orbital eccentricities}

With seventeen confirmed and one likely binary systems, the binary
frequency of our sample is 77$-$82$\pm$10\%, taking Poisson sampling
errors into account, but $\approx$ 20\% of the stars appear 
to be single. Thus, the frequency of spectroscopic binaries among
CEMP-$s$ stars is clearly much higher than among both Population I and
II giants in the field \citep{mermio2007, carney2003}, but as discussed 
in Sect. \ref{singles}, it is probably {\it not quite} the 100\% 
surmised by \citet{lucatello2005}. That $\approx$ 20\% of the stars
remain single suggests that, like the CEMP-no stars of Paper~II, 
they did not receive their high carbon and
$s$-process-element abundances via mass transfer from a former AGB companion;
another scenario must be invoked (see Sect. \ref{singlestars}).   

\begin{figure}
\resizebox{\hsize}{!}{\includegraphics{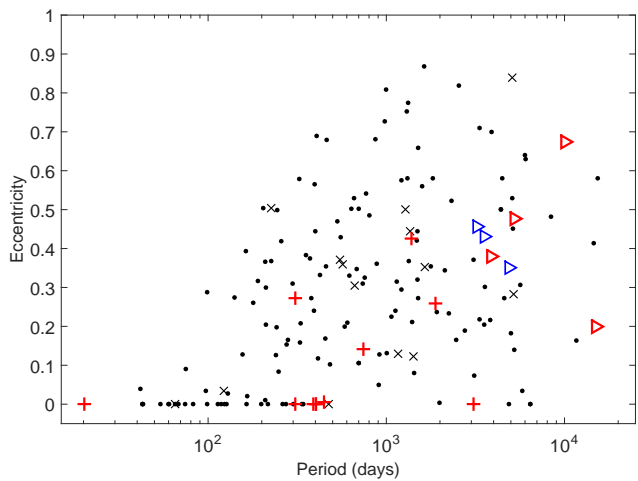}}
\caption{Period$-$eccentricity diagram for the binary systems in our sample. 
Red and blue symbols denote CEMP-$s$ and CEMP-$r/s$ stars, respectively. 
Systems with periods $P < 3,000$ days are plotted as right-pointing triangles 
to indicate that the orbital parameters are still uncertain -- their periods, 
in particular, may well be even longer. {\it Black dots and crosses:} Comparison 
sample of 141 giant binary members of Galactic open clusters of all ages
\citep{mermio2007,mathieu1990} and 16 metal-poor and C-normal binaries from 
\citet{carney2003}.}  
\label{fig:Pe}
\end{figure}

Figure~\ref{fig:Pe} shows our binary systems (Table \ref{tbl-3}) in the 
period-eccentricity diagram (red plus signs and red and blue triangles),
along with comparison samples of 141 giant binary members of open clusters of
all ages by \citet{mermio2007} and \citet{ mathieu1990} (black dots),
and 16 metal-poor binaries from \citet{carney2003} (black crosses). 
\object{HE~0959$-$1424} has been included in the plot with a fictitious 
period of 15,000 days to indicate that its period is likely very long, but 
presently unknown. The three CEMP-$r/s$ stars (HE~0017$+$0055, HE~0039$-$2635, 
and LP~625$-$44; blue triangles) have similar eccentric orbits with 
long periods (of the order a decade or more), and are among the 
longest-period stars in our sample (Table \ref{tbl-3}). However, the 
present sample is too small to claim any difference in binary characteristics 
between CEMP-$r/s$ and CEMP-$s$ stars (see further discussion in Sect. \ref{cemprs}).

The period-eccentricity diagram shown in Figure~\ref{fig:Pe} exhibits the
expected overall feature \citep{jorissen1998,jorissen2015b} of (near-) 
circular orbits up to a cutoff that depends on the age of the systems, but 
can be estimated at $\approx$ 200 days for Population I and Population II 
giants of all ages. This is commonly ascribed to tidal circularisation 
of the orbits, which is more advanced the older the star (the few 
moderately eccentric shorter-period binaries are probably younger and/or 
more metal-poor stars in uncertain stages of evolution). 

The CEMP-$s$ and CEMP-$r/s$ binaries shown in Figure~\ref{fig:Pe} indicate a
cutoff period in the range 500$-$700 days, suggestive of larger (AGB) 
secondaries. Two systems, HE~1046$-$1352 and HE~1523$-$1155, are exceptions 
to this general trend. HE~1046$-$1352, with a period of only 20 days, does 
have the expected circular orbit, but the calculated Roche-lobe radii for an 
average-mass white dwarf secondary are too small to accommodate even a normal 
giant star, let alone an AGB star of $\approx200$~R$_{\sun}$. HE~1046$-$1352 
could be a misclassified dwarf or the result of a common-envelope phase of 
evolution, similar to the even shorter-period system \object{HE~0024$-$2523} 
discussed by \citet{lucatello2003} on the basis of its high spectroscopic 
log $g$, remarkable abundance pattern, and high rotation. 

HE~1523$-$1155 deviates from expectation in the opposite sense, in that 
it has a highly significant eccentricity of $e=0.3$, yet has a 
shorter period (309 days) than the $\approx$ 600-day cutoff defined 
by the several systems with circular orbits and periods near 500 days. 
As noted above, the period is observationally strongly constrained, so 
any additional observations could only confirm the currently derived 
eccentricity or lead to an even higher value. One might suspect whether 
HE~1523$-$1155 could also be a dwarf. However, the spectroscopic 
$\log$ $g$ = 1.6 derived by \citet{aoki2007} from a Subaru spectrum with 
$R \approx 50,000$ indicates a giant classification for this star. 

\section{Formation of CEMP-$s$ stars via mass transfer}

In analogy with the higher-metallicity CH and Ba stars, and based on recent 
detailed models for them (e.g., \citealt{abate2015a} and references therein),
CEMP-$s$ stars can be produced through local mass transfer from an AGB binary
companion. However, challenges for this scenario remain, such as: $(i)$ Can
models account for their observed frequencies at low metallicity? (previous
attempts fall well short of matching the observations, but progress is being
made, see \citealt{abate2015b}); $(ii)$: Can the models reproduce the
observed distribution of periods and (final) separations of the binary
components?, and $(iii)$: Can the models account for the detailed abundance
patterns of CEMP-$s$ (and CEMP-$r/s$) stars? Final resolution of these
questions will require substantially larger samples of CEMP-$s$ (and CEMP-$r/s$)
stars with precise long-term RV-monitoring and high-SNR, high-resolution
spectra. Below we speculate on what can be inferred from the present
sample. 

\subsection{Mass-transfer mechanisms}

The transfer of mass from an AGB companion star onto the low-mass star we 
observe today as a CEMP-$s$ star can happen in three more-or-less efficient 
ways: Roche-lobe overflow (RLOF), wind transfer, or wind-assisted Roche-lobe 
overflow (WRLOF). 

Roche-lobe overflow occurs when a binary component 
expands beyond its Roche-lobe radius. Mass can then be transferred to
the companion through the inner Lagrangian point (L1). This is the most
efficient way to transfer mass in a binary system, but for RLOF to operate, 
the separation of the two stars in the system must be relatively small; if the
stars are too far apart, neither of them will ever fill its Roche lobe. On the
other hand, if the stars are too close, they will enter a common-envelope
phase. If a binary system undergoes RLOF, this will circularise the orbit very
effectively; even more so if it enters a common-envelope phase. RLOF from
  stars with large convective envelopes, such as AGB stars, is generally
  believed to be unstable and often develops into a common-envelope phase
  \citep{kopal1959,paczynski1965,paczynski1976}. Mass transfer during the
  common envelope phase is very inefficient and usually considered to be
  negligible \citep{ricker2008}; therefore, RLOF is generally not considered a
  possible formation mechanism for CEMP-s stars.  

From Table~\ref{tbl-3}, it appears that the calculated secondary
Roche-lobe radii for our binary systems are only large enough to
accommodate an AGB star of $\approx200$~R$_{\sun}$ in systems with orbital
periods $P \ga 1,000$ days and massive present-day WD companions. This, 
along with the obstacles of common-envelope evolution as mentioned above,
would seem to rule out the direct RLOF of mass transfer in shorter-period
binaries and favour the WRLOF mode discussed below. 

Another method for mass transfer in a binary system is wind transfer,
where the low-mass star is exposed to the wind of the AGB star, and can
thereby accrete mass \citep{bondihoyle1944}. This Bondi-Hoyle-Lyttleton
(BHL) model of wind mass transfer assumes that the wind of the AGB star 
does not interact with the orbit of the accreting star, which requires 
the wind velocity ($v_{wind}$) to be much higher than the orbital velocity 
($v_{orbit}$); thus, this type of mass transfer is an option only if 
$v_{wind} >> v_{orbit}$. 

This is not always the case for wide binary systems, however. If 
$P \approx 10^4$ days, the orbital velocity is $\approx$ 10
km~s$^{-1}$, and the wind from AGB stars can have velocities of 5-30
km~s$^{-1}$ \citep{vassiliadis1993}. \citet{boffin1988} explored the
possibility of creating Ba stars (the higher-metallicity equivalent of
CEMP-$s$ stars) via wind transfer in detached binary systems using the
BHL wind-accretion scenario, and they concluded that the Ba stars
could indeed have been formed in this way. This type of mass transfer allows
orbital periods to remain long and orbits to not be circularised. 

A third option is the WRLOF mass-transfer mechanism, which can occur in
systems where the wind of the donor star is gravitationally confined to
the Roche-lobe of the accreting star. The wind can then be focussed
towards the orbital plane of the binary system and transferred to the
secondary through the L1 point. This type of transfer can be
significantly more efficient than BHL wind transfer
\citep{mohamed2007}. The WRLOF mechanism facilitates mass transfer in
binary systems that are too wide for mass transfer via conventional
RLOF. The majority of the material that is not accreted by the secondary
is lost through the outer Lagrangian points (L2 and L3), and thus
carries away angular momentum from the system, shrinking the orbit
\citep{abate2013}.

\citet{abate2015a} attempted to reproduce the chemical abundance
pattern and orbital properties of 15 known CEMP-$s$ binary systems. To
do this, they compared two models: (1) a WRLOF model with angular momentum
loss calculated assuming a spherical symmetric wind, and (2) an enhanced
BHL wind-transfer model with efficient angular momentum loss. 

Depending on the specific model adopted for the angular-momentum loss, the 
binary system can widen or shrink in response. Models with WRLOF and a
spherically-symmetric wind widen the orbit as mass is transferred, i.e., small
initial periods and consequently small secondary masses are required. 
Otherwise, they would overflow their Roche lobes, which would lead to a 
common-envelope phase with no accretion onto the companion. Because the
WRLOF is not very efficient for small separations, only small amounts of
material is transferred, and this model fails to match the measured
abundances of the CEMP-$s$ stars. 

In contrast, the BHL wind-transfer model with enhanced angular-momentum
loss predicts high accretion efficiency (also for systems with
relatively short periods), and the enhanced angular-momentum loss ensures
that the systems always shrink in response to the mass loss. Thus, even
systems that are initially very wide can end up being relatively close
when the primary has become a white dwarf. 

The majority of their CEMP-$s$ binary systems were best reproduced by
the second model -- a model the authors doubted is realistic, but which
emphasized the need for very efficient mass transfer to create the
abundance patterns observed in the CEMP-$s$ stars. The systems modelled
by \citet{abate2015a} have periods ranging between 3.33 and 4,280 days,
and they find initial periods of 1,170 to 130,000 days for these
systems. 

Hence, \citet{abate2015a} conclude that, to reproduce the observations,
``... it is generally necessary that the modelled binary systems lose
efficiently angular momentum and transfer mass with high accretion
efficiency.'' In some systems, a common-envelope
phase is even needed to shrink the orbits to their current size. This is
particularly true for the binary system \object{HE~0024$-$2523}, with a
period of only 3.41 days \citep{lucatello2003}.

\subsection{Nucleosynthesis challenges}

Aside from the challenges of efficiently transferring mass in a binary
system, most AGB models also face difficulty in reproducing the detailed
elemental-abundance patterns found in CEMP-$s$ stars. For instance,
 \citet{bisterzo2012} fitted the abundance patterns of 94 individual CEMP-$s$
 and CEMP-$r/s$ stars with yields from AGB-star models. The outcome of that
 exercise demonstrated that the models have problems reproducing the C and N
 abundances, as well as the $^{12}$C/$^{13}$C ratios reported for CEMP
 stars. Carbon is generally over-produced in their models. Combined with the
 observed low $^{12}$C/$^{13}$C ratios (see, e.g., \citealt{ryan2005} and
\citealt{hansen2015a}, and references therein), this points toward a
large amount of mixing not being included in their models. Recent
work by \citet{abate2015a} and \citet{abate2015b}, employing models of
AGB nucleosynthesis production from \citet{karakas2010}, reaches better
agreement for some stars, while for others they have the same problem as
\citet{bisterzo2012}.

In addition, predictions of the $s$-process-element distributions
present difficulties. The AGB models of \citet{bisterzo2012} predict
roughly similar abundances (within 0.3 dex) of the first-peak (Sr, Y,
Zr) and second-peak elements (Ba, La, Ce, Pr, and Nd), while the derived
abundances for CEMP-$s$ stars exhibit an internal spread in these
elements of more than 0.5 dex. Barium is often found to be more enhanced
than the other second-peak elements.   

\citet{placco2013} attempted to match the measured abundances of CEMP-$s$ stars
(without scaling the pattern to the Ba abundance) with yields from the
AGB models of \citet{karakas2010}, taking into account the dilution
across the binary system. In this case, the elements from the first
$s$-process peak were under-produced, while Ba and Pb were over-produced
by the model. Alternatively, \citet{hansen2015a} compared yields from the
F.R.U.I.T.Y database of \citet{cristallo2009,cristallo2011} to the measured
abundance pattern of two CEMP-$s$ stars. These models failed to reproduce the
large amounts of C and N detected in the stars, and also have problems
matching the abundances derived for the light neutron-capture elements (Sr, Y,
and Zr) of the stars. 

Near-ultraviolet (NUV) observations of CEMP-$s$ stars also help constrain the
theoretical models. \citet{placco2015} studied two bright CEMP-$s$ stars with
HST/STIS spectra, which allowed them to determine abundances for elements
such as Ge, Nb, Mo, Lu, Pt, and Au. They also matched the abundance pattern 
with the latest models from \citet{abate2015a}, which take into account not 
only the AGB evolution, but also the dynamics of the binary system.
Even though NUV abundance determinations for CEMP stars are challenging to
obtain, they are important inputs for such model-matching exercises.

\subsection{Dilution of the accreted material}

The accreted material also mixes with interior material from the accreting
stars via thermohaline mixing \citep{stancliffe2007}, which will result in 
dilution of the accreted material and a change in its abundance pattern. 
Attempts to model this effect have been made, but
the degree to which the transferred material is diluted on the surface of the
receiving star is still very poorly constrained \citep{stancliffe2007,
stancliffe2008}. 

The accreted material will also be mixed with the material of the accreting 
star during any dredge-up episodes that occur as the low-mass CEMP-$s$ star
evolves past the main-sequence stage. This is the case for
some of the stars modelled by \citet{abate2015a}, which need to have
even more mass transferred to account for the mixing occurring during first
dredge-up in order to match the abundance pattern of the observed CEMP-$s$
star. In addition, \citet{richard2002} argue that the degree of dilution 
depends on the chemical composition of the accreted material. These different
mixing effects change the composition of the surface of the CEMP-$s$
star, and introduce yet another source of uncertainty into the
mass-transfer models.

\section{The CEMP-$r/s$ stars}
\label{cemprs}
Europium has been detected in four stars of our sample (see Table \ref{tbl-Eu}), 
indicating an enhancement in $r$-process as well as $s$-process elements. 
One of these, CS~30301$-$015, we have now classified as a CEMP-$s$ star; it is
also a single star. The other three, HE~0017$+$0055, HE~0039$-$2635, and
LP~625$-$44, are classified as CEMP-$r/s$ stars and are long-period binaries
with orbital properties that are indistinguishable from those of CEMP-$s$
stars. 

Based on Ba and Eu abundances for CEMP-$s$ and CEMP-$r/s$ stars, 
\citet{lugaro2012} argued that the abundance patterns found for the
stars cannot come from the same single AGB star source,  essentially 
opposite to the claim by \citet{allen2012} that the astrophysical origin 
of the CEMP-$s$ and CEMP-$r/s$ stars is one and the same. 

It has also been speculated that the CEMP-$r/s$ stars may exhibit the
chemical signature of a different origin, the intermediate
neutron-capture ($i$-)process. The $i$-process, operating at
neutron fluxes between the $s$- and $r$-process, was first proposed
by \citet{cowan1977} to occur in evolved red giants. This was supported
by \citet{masseron2010}, who argued that CEMP-$r/s$ stars could be the
result of mass transfer from more massive AGB stars, where the neutron
source $^{22}$Ne($\alpha$,n) $^{25}$ is active during the thermal
pulses. This source is predicted to produce sufficiently high
neutron exposures to not only synthesize elements traditionally
associated with the $s$-process, but also elements such as Eu, which are
normally associated with the $r$-process \citep{goriely2005}. More
recently, \citet{bertolli2013} also found that the abundance pattern of
CEMP-$r/s$ stars could be matched with the yields from the $i$-process,
thought to occur in high-mass ``super-AGB'' stars. 

We have searched the literature for CEMP stars with $r$- and
$s$-process element abundances such that they would qualify as
CEMP-$r/s$ stars, and with published radial-velocity data. Nine such stars
show radial-velocity variations, but only three have published orbital
elements (see Table \ref{tbl-cemprs}); five show no variation. However, the
information on their stage of evolution (essentially a spectroscopic log $g$)
is ambiguous. 

Accordingly, no significant new information can be derived from a
$P$-$e$ diagram such as that in Figure \ref{fig:Pe}, except that the
orbital periods of the certified giant CEMP-$r/s$ stars are long and
their eccentricities are significant, but moderate -- very similar to
those of the rest of the CEMP-$r/s$ stars. Larger samples of CEMP-$r/s$
stars with detailed information on binary orbital parameters and a
more complete inventory of heavy neutron-capture elements are clearly
required for further progress. 

\begin{table}
\caption{Orbital elements for CEMP-$r/s$ binary stars from the literature}
\label{tbl-cemprs}
\centering
\resizebox{0.48\textwidth}{!}{
\begin{tabular}{lrrl}
\hline\hline
Stellar ID & Period & $e$ & Ref\\
\hline
\object{CS~22948$-$027}&  427 & 0.02 & \citet{barbuy2005}\\
\object{CS~29497$-$030}&  346 & 0.30 & \citet{preston2000}\\
\object{HD~224959}     & 1273 & 0.18 & \citet{mcclure1990}\\
\hline
\end{tabular}}
\end{table}

\section{Composition of the single stars}
\label{singlestars}

Four stars in our sample exhibit constant velocities over our observing period, 
and are judged to be single (see Fig. \ref{fig:constant} and Sect. \ref{singles}): 
HE~0206$-$1916, HE~1045$+$0226, HE~2330$-$0555, and CS~30301$-$015. The
atmospheric parameters for these single stars are listed in Table
\ref{tbl-param}. They are all giants, which may have experienced a first
dredge-up (FDU) episode. During this FDU, H, He, and products of the CN cycle
are mixed to the surface, resulting in a small decrease in $\mathrm{[C/Fe]}$ along
with an increase in $\mathrm{[N/Fe]}$ at the surface of the star (the
precise abundance changes depend on the mass and metallicity of the star
and the extent of the FDU, see e.g. \citealt{karakas2014}). Thus, no great
change of their surface abundances is expected to have occurred, and a
different channel must be invoked to account for the origin of the carbon and
$s$-process-element enhancements found in these stars. 

\begin{figure}
\resizebox{\hsize}{!}{\includegraphics{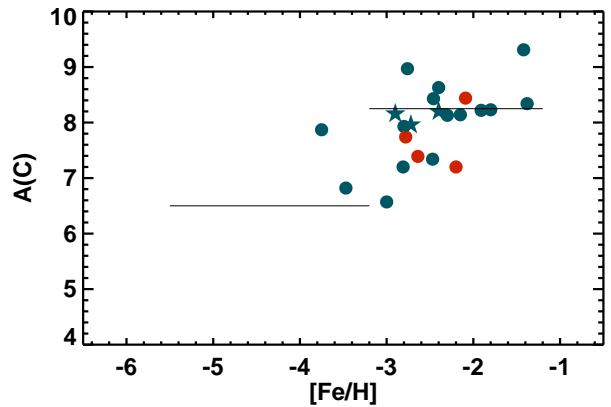}}
\caption{Absolute C abundances for our programme stars.  Dots represent
  CEMP-$s$ stars, star symbols CEMP-$r/s$ stars; red symbols show single stars, 
  blue symbols certified binaries. The two C-bands first noted by 
  \citet{spite2013} are indicated by horizontal lines.} 
\label{fig:Cband}
\end{figure}

\begin{figure}
\resizebox{\hsize}{!}{\includegraphics{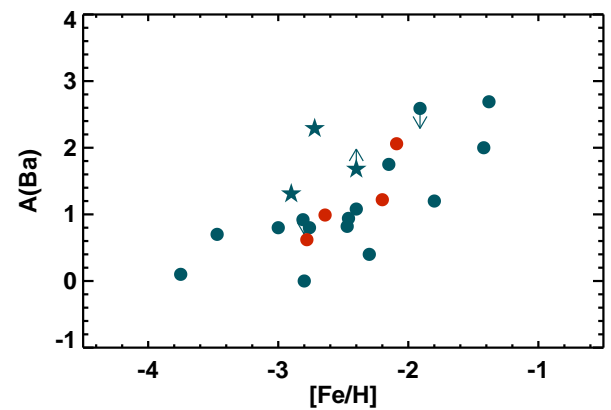}}
\caption{Absolute Ba abundances for our programme stars. Dots represent
  CEMP-$s$ stars, star symbols CEMP-$r/s$ stars; red symbols show single stars, 
  blue symbols are definite binaries.}
\label{fig:Ba}
\end{figure}

To explore this possibility, we have plotted the abundance information we have
in common for all of the stars, namely the C, Fe, and Ba abundances; see Figures
\ref{fig:Cband} and \ref{fig:Ba}. The single stars (red dots) exhibit no
clear difference in abundance signatures for these elements as compared with
the binary stars (blue dots and stars). More detailed abundance analyses, spanning a greater range of elements than is available at present, may reveal
differences in the abundance patterns of the single stars vs. those found in
binary systems, which could help constrain the likely formation scenario of
these stars.

\begin{table}
\caption{Stellar parameters for single stars}
\label{tbl-param}
\centering
\resizebox{0.48\textwidth}{!}{
\begin{tabular}{lcccl}
\hline\hline
Stellar ID & T$_{\rm eff} (K)$ & $\log g$ & [Fe/H] & Ref \\
\hline
HE~0206$-$1916 & 5200 & 2.7& $-$2.09 & \citet{aoki2007}\\
HE~1045$+$0226 & 5077 & 2.2& $-$2.20 & \citet{cohen2013}\\
HE~2330$-$0555 & 4900 & 1.7& $-$2.78 & \citet{aoki2007}\\
CS~30301$-$015 & 4750 & 0.8& $-$2.64 & \citet{aoki2002b}\\
\hline
\end{tabular}}
\end{table}

\subsection{Potential sources of carbon and $s$-process elements in the early Universe}

The carbon excess detected in the single CEMP-$s$ stars may have a
similar origin as the carbon excess seen in the CEMP-no stars, which
is believed to be due to either winds from the so-called ``spinstars'' --
massive, rapidly-rotating, metal-free stars \citep{meynet2006,
hirschi2007, maeder2015}, or to faint supernovae with mixing and 
fallback as suggested by \citet{umeda2003}. 

Spinstars are potentially also a main contributor to 
$s$-process-element abundances in the early Universe. It was first suggested by
\citet{pignatari2008} that rotation-induced mixing in these stars
enhances the production of $s$-process elements via the weak $s$-process. 
Later, \citet{frischknecht2012,frischknecht2015} explored the yields of the
weak $s$-process in spinstars. They found that mainly light
$s$-process elements (Sr, Y, Zr) are produced in this environment,
usually with $\mathrm{[Sr/Ba]} > 0$; only in extreme cases does the
production run all the way through to Pb. 

Abundances of Sr, Ba, and Pb have only been reported in the literature
for one of our single stars, CS~30301$-$015; with $\mathrm{[Sr/Ba]} =
-1.15$ and $\mathrm{[Pb/Fe]} = +1.7$ \citep{aoki2002b}, it does not
match the predictions from \citet{frischknecht2012}. On the other hand,
detailed models of the abundance patterns generated by spinstars have
not yet matured to the point where meaningful star-by-star comparisons
can be carried out. Such models might help to identify the progenitors
of the single CEMP-$s$ stars.

\subsection{Carbon bands}

\citet{spite2013} first suggested the existence of two bands in the absolute
carbon abundances of CEMP stars -- a high-C band at $A(\rm C) \approx 8.25$
and a low-C band at $A(\rm C) \approx 6.5$. The existence of these two
bands for CEMP stars has been confirmed with larger samples by
\citet{bonifacio2015} and \citet{hansen2015a}. The high-C band is
primarily populated by relatively more metal-rich stars
($\mathrm{[Fe/H]}\gtrsim -3$) of the CEMP-$s$ sub-class, while the low-C
band is primarily populated by more metal-poor stars of the CEMP-no
sub-class. While this general behaviour is supported by the CEMP-no
(Figure 4 of Paper~II), CEMP-$s$, and CEMP-$r/s$ (Figure~\ref{fig:Cband}) 
stars in our programme, there are clear exceptions. For instance, in
Figure~\ref{fig:Cband} a substantial number of stars lie in the transition
zone between the two carbon bands, and two stars, classified as CEMP-$s$
stars on the basis of their $\mathrm{[Ba/Fe]}$ abundance ratios, are
nevertheless found close to the low-C band, which is thought to be
predominantly associated with CEMP-no stars. 

It is remarkable that these stars, HE~0430$-$1609 and HE~2312$-$0758,
are two of the three lowest-metallicity stars in our sample. Both 
stars are confirmed binaries with long-period 
orbits. The presence of these stars on the low C-band could be due to
less-efficient mass transfer in these systems. Perhaps no mass-transfer
event has occurred in these systems at all, and the CEMP-$s$ stars
received their C and $s$-process-element excesses from sources similar
to the single CEMP-$s$ stars. The lowest metallicity CEMP-$s$ star in
our programme, HE~0002$-$1037, lies close to the high-C band, similar to
the star HE~0959$-$1424, which is at the high-metallicity limit for
stars in our programme and exhibits a carbon abundance that is well
above the high-C band. 

The single stars in our sample form a heterogeneous group. Three of the
stars, HE~1045$-$0226, CS~30301$-$015, and HE~2330$-$0555, lie in the
transition zone between the carbon bands, while the fourth,
HE~0206$-$1916, lies on the high-C band. 

There clearly remains much work in order to assess the meaning, and
possible astrophysical significance, of the suggested carbon bands.
Whether this turns out to be a primary behaviour, with deviations that
occur for individual stars depending on their precise nucleosynthesis
histories, or is correlated with some other factor related to the
production and/or dilution of the carbon due to mixing, awaits larger
samples with available RV-monitoring and high-SNR spectroscopy.

\section{Conclusions}

Our systematic and precise long-term radial-velocity monitoring programme of 
22 CEMP-$s$ and CEMP-$r/s$ stars has led to a high frequency, 82$\pm$10\%, of
detected binaries in the sample, confirming that the characteristic abundance
pattern of the CEMP-$s$ stars is coupled to the binary nature of most of these
stars. Indeed, it has been taken for granted up to now that {\it all} CEMP-$s$
stars originated from mass transfer in a binary system, a basic assumption which
underpins all previous theoretical modelling of these stars. 

Four of our programme stars (18\%) do, however, appear to be truly single and contradict this 
basic assumption -- as do in fact the entire group of CEMP-no stars discussed 
in Paper~II. This underscores the need for an efficient,
distant source of $s$-process elements in the early Universe, whether that
source might be spinstars or another class of progenitor. Future detailed
abundance analyses of these single stars may reveal signatures in their 
abundance patterns that will help to constrain the nature of the early C and
$s$-process-element production sites and the processes by which these fresh
elements were implanted in the natal clouds of the observed CEMP-$s$ stars.  

For the binary CEMP-$s$ stars, we highlight a number of respects,
notably the detailed treatment of the physics of the mass transfer process, in
which current models still fail to reproduce their observed orbital parameters
and elemental abundance patterns.   

The three CEMP-$r/s$ stars in our sample are all long-period binaries, 
hinting that the origin of these stars is similar to that of the CEMP-$s$ 
stars themselves. However, the available data are much too sparse to be 
conclusive on such key points as whether their excesses of C and 
neutron-capture elements stem from the same nuclear source(s), and whether 
the latter are due to a special form of the $s$-process or to the
$i$-process, which has been speculated to operate in the so-called 
``super-AGB'' stars. 

Larger samples of binary CEMP-$s$ and CEMP-$r/s$ stars with well-determined 
orbits, evolutionary status, and detailed abundance patterns are required 
for further refinement of the theoretical models for the origin of these 
stars. The accurate trigonometric parallaxes and accompanying precise 
multi-epoch photometry for many CEMP stars in the field to be expected from 
the ESA space mission Gaia will soon provide further crucial information on
the stage of evolution and pulsational properties of these stars.  

\noindent {\bf Note added in proof:} After this paper was accepted, we were
able to obtain a final FIES spectrum of \object{HE~0430$-$1609}. Remarkably, 
this velocity still fits two very different orbits with the same rms error 
of only 40 m s$^{-1}$. We retain the longer-period circular orbit as the most 
likely and have updated the tables and figures accordingly. 

\begin{acknowledgements}

This paper is based on observations made with the Nordic Optical
Telescope, operated by the Nordic Optical Telescope Scientific
Association at the Observatorio del Roque de los Muchachos, La Palma,
Spain, of the Instituto de Astrof{\'i}sica de Canarias. We thank numerous 
NOT staff members and students for readily and very efficiently 
obtaining most of the many observations for us in service mode. We also 
thank Carlo Abate for clarifying the response of his modelled binary systems
to the different modes of angular-momentum loss. Finally, we thank the 
anonymous referee for a very careful and incisive report that led to 
substantive clarifications, revisions and improvements in the paper. \\

This work was supported by Sonderforschungsbereich SFB
881 ``The Milky Way System'' (subproject A4) of the German Research
Foundation (DFG). J.A. and B.N. gratefully acknowledge financial support
from the Danish Natural Science Research Council and the Carlsberg
Foundation. T.C.B., V.M.P., and J.Y. acknowledge partial support for
this work from grants PHY 08-22648; Physics Frontier Center/{}Joint
Institute or Nuclear Astrophysics (JINA), and PHY 14-30152; Physics
Frontier Center/{}JINA Center for the Evolution of the Elements
(JINA-CEE), awarded by the US National Science Foundation.

\end{acknowledgements}


\clearpage

\begin{appendix}
\section{Radial Velocities Measured for the Programme Stars}

\begin{table}[h]                   
\caption{HE~0002$-$1037}         
\centering                       
\begin{tabular}{lcc}
\hline\hline
HJD & RV & RV$_{err}$  \\
    & km~s$^{-1}$ & km~s$^{-1}$\\
\hline
2456191.52615  & $-$36.229 & 0.045\\
2456241.45815  & $-$33.883 & 0.184\\
2456530.70894  & $-$24.359 & 0.040\\
2456574.57567  & $-$27.581 & 0.044\\
2456647.45601  & $-$35.052 & 0.081\\
2456652.37785  & $-$35.635 & 0.053\\
2456888.57469  & $-$38.668 & 0.061\\
2456956.43565  & $-$34.844 & 0.076\\
2457225.65699  & $-$22.941 & 0.065\\
2457257.63749  & $-$23.761 & 0.055\\
\hline
\end{tabular}           
\end{table}

\begin{table}[h]                  
\caption{HE~0017$+$0055}         
\centering                       
\begin{tabular}{lcc}
\hline\hline
HJD & RV & RV$_{err}$  \\
    & km~s$^{-1}$ & km~s$^{-1}$\\
\hline
2454314.67018 & $-$78.622 & 0.007\\
2454338.64193 & $-$78.944 & 0.007\\
2454373.62240 & $-$78.691 & 0.009\\
2454396.53706 & $-$78.583 & 0.007\\
2454406.59662 & $-$78.505 & 0.012\\
2454480.38681 & $-$78.183 & 0.012\\
2454793.48462 & $-$79.098 & 0.018\\
2454820.33862 & $-$78.830 & 0.017\\
2455059.73646 & $-$79.922 & 0.008\\
2455149.47303 & $-$79.647 & 0.010\\
2455207.34980 & $-$79.391 & 0.015\\
2455415.60806 & $-$80.439 & 0.009\\
2455439.59141 & $-$80.778 & 0.010\\
2455503.40860 & $-$80.330 & 0.023\\
2455738.73436 & $-$80.148 & 0.008\\
2455776.68215 & $-$80.675 & 0.008\\
2455821.57659 & $-$80.870 & 0.009\\
2455944.32531 & $-$80.429 & 0.011\\
2456139.71293 & $-$80.857 & 0.010\\
2456241.39102 & $-$81.420 & 0.018\\
2456545.62665 & $-$81.474 & 0.011\\
2456652.41797 & $-$81.951 & 0.015\\
2456686.32089 & $-$81.538 & 0.015\\
2456840.71823 & $-$80.866 & 0.028\\
2456888.54207 & $-$81.261 & 0.013\\
2456917.59811 & $-$81.398 & 0.014\\
2456987.38131 & $-$81.885 & 0.009\\
2457257.57737 & $-$81.396 & 0.010\\
\hline
\end{tabular}           
\end{table}

\begin{table}[h]                   
\caption{HE~0111$-$1346}         
\centering                       
\begin{tabular}{lcc}
\hline\hline
HJD & RV & RV$_{err}$  \\
    & km~s$^{-1}$ & km~s$^{-1}$\\
\hline
2456213.65944  & $+$50.431 & 0.022\\
2456531.67451  & $+$39.604 & 0.015\\
2456534.72823  & $+$40.177 & 0.024\\
2456574.62161  & $+$47.100 & 0.015\\
2456579.62258  & $+$47.786 & 0.020\\
2456652.40265  & $+$49.094 & 0.021\\
2456893.65246  & $+$31.577 & 0.018\\
2456917.61673  & $+$36.046 & 0.019\\
2457257.65967  & $+$26.462 & 0.026\\
\hline
\end{tabular}           
\end{table}

\begin{table}[h]                   
\caption{HE~0151$-$0341}         
\centering                       
\begin{tabular}{lcc}
\hline\hline
HJD & RV & RV$_{err}$  \\
    & km~s$^{-1}$ & km~s$^{-1}$\\
\hline
2456213.70382 & $-$37.298 & 0.070\\
2456307.32814 & $-$25.696 & 0.064\\
2456518.73462 & $-$47.217 & 0.041\\
2456545.65829 & $-$42.753 & 0.047\\
2456574.65686 & $-$36.767 & 0.030\\
2456603.50990 & $-$31.030 & 0.042\\
2456647.48211 & $-$25.892 & 0.065\\
2456652.47616 & $-$25.703 & 0.072\\
2456685.39340 & $-$26.739 & 0.065\\
2456893.72033 & $-$44.812 & 0.051\\
2457225.72100 & $-$48.629 & 0.065\\
\hline
\end{tabular}           
\end{table}            

\begin{table}                   
\caption{HE~0206$-$1916}         
\centering                       
\begin{tabular}{lcc}
\hline\hline
HJD & RV & RV$_{err}$  \\
    & km~s$^{-1}$ & km~s$^{-1}$\\
\hline
2456213.68126  & $-$199.551 & 0.047\\
2456307.37509  & $-$199.446 & 0.052\\
2456529.69669  & $-$199.422 & 0.024\\
2456546.73298  & $-$199.723 & 0.060\\
2456603.53471  & $-$199.651 & 0.041\\
2456893.67497  & $-$199.471 & 0.033\\
2456987.43608  & $-$199.626 & 0.061\\
2457018.49730  & $-$199.580 & 0.073\\
2457257.68120  & $-$199.356 & 0.037\\

\hline
\end{tabular}           
\end{table}            

\begin{table}                   
\caption{HE~0319$-$0215}         
\centering                       
\begin{tabular}{lcc}
\hline\hline
HJD & RV & RV$_{err}$  \\
    & km~s$^{-1}$ & km~s$^{-1}$\\
\hline
2454780.64369 & $-$229.933 & 0.029\\
2455126.64403 & $-$227.058 & 0.024\\
2455149.51200 & $-$226.974 & 0.030\\
2455207.38599 & $-$226.673 & 0.017\\
2455415.70963 & $-$224.274 & 0.023\\
2455439.67958 & $-$224.389 & 0.019\\
2455478.61056 & $-$224.219 & 0.022\\
2455620.34838 & $-$223.797 & 0.026\\
2455821.65016 & $-$222.893 & 0.012\\
2455882.56766 & $-$223.207 & 0.020\\
2455915.54130 & $-$223.575 & 0.117\\
2456191.63308 & $-$223.890 & 0.017\\
2456528.72760 & $-$226.084 & 0.028\\
2456578.64692 & $-$226.565 & 0.018\\
2456956.69477 & $-$229.280 & 0.085\\
2456987.52074 & $-$229.692 & 0.063\\
\hline
\end{tabular}           
\end{table}            

\begin{table}                   
\caption{HE~0430$-$1609}         
\centering                       
\begin{tabular}{lcc}
\hline\hline
HJD & RV & RV$_{err}$  \\
    & km~s$^{-1}$ & km~s$^{-1}$\\
\hline
2456209.69059   & $+$228.767 & 0.025\\
2456214.65424   & $+$228.731 & 0.021\\
2456529.71626   & $+$230.277 & 0.020\\
2456545.73032   & $+$230.365 & 0.020\\
2456574.68459   & $+$230.530 & 0.015\\
2456685.46062   & $+$231.151 & 0.018\\
2456722.36199   & $+$231.359 & 0.048\\
2456893.69735   & $+$232.234 & 0.035\\
2456917.63747   & $+$232.325 & 0.033\\
2456956.76841   & $+$232.672 & 0.116\\
2456986.65656   & $+$232.743 & 0.022\\
2457076.35811   & $+$233.154 & 0.086\\
2457092.34638   & $+$233.248 & 0.025\\
2457110.35827   & $+$233.315 & 0.050\\
2457257.71794   & $+$233.909 & 0.029\\
2457393.47107   & $+$234.389 & 0.037\\
\hline                    
\end{tabular}             
\end{table}            

\begin{table}                   
\caption{HE~0441$-$0652}         
\centering                       
\begin{tabular}{lcc}
\hline\hline
HJD & RV & RV$_{err}$  \\
    & km~s$^{-1}$ & km~s$^{-1}$\\
\hline
2454705.73510 &  $-$39.600 & 0.154\\
2455149.58120 &  $-$33.053 & 0.639\\
2455176.49676 &  $-$31.568 & 0.372\\
2455207.53305 &  $-$30.860 & 0.149\\
2455439.70545 &  $-$29.741 & 0.163\\
2455478.64746 &  $-$29.967 & 0.081\\
2455531.57019 &  $-$29.901 & 0.271\\
2455821.67913 &  $-$28.773 & 0.130\\
2455915.58170 &  $-$27.892 & 0.810\\
2456005.34969 &  $-$29.417 & 0.507\\
2456191.66229 &  $-$29.167 & 0.183\\
2456214.67479 &  $-$29.727 & 0.170\\
2456574.73303 &  $-$30.053 & 0.135\\
2456685.48885 &  $-$29.826 & 0.417\\
2456987.46556 &  $-$30.047 & 0.298\\
2457076.43392 &  $-$30.762 & 1.056\\
\hline
\end{tabular}           
\end{table}            

\begin{table}                   
\caption{HE~0507$-$1430}         
\centering                       
\begin{tabular}{lcc}
\hline\hline
HJD & RV & RV$_{err}$  \\
    & km~s$^{-1}$ & km~s$^{-1}$\\
\hline
2455149.65141 &  $+$53.744 & 0.017\\
2455176.59963 &  $+$53.614 & 0.027\\
2455207.45478 &  $+$51.698 & 0.016\\
2455232.47668 &  $+$48.880 & 0.016\\
2455478.68157 &  $+$40.340 & 0.029\\
2455821.73671 &  $+$32.139 & 0.018\\
2455882.66528 &  $+$34.746 & 0.032\\
2455944.52431 &  $+$43.196 & 0.044\\
2455971.41362 &  $+$47.094 & 0.037\\
2456005.41341 &  $+$51.227 & 0.077\\
2456213.73374 &  $+$36.148 & 0.030\\
\hline
\end{tabular}           
\end{table}            

\begin{table}                   
\caption{HE~0507$-$1653}         
\centering                       
\begin{tabular}{lcc}
\hline\hline
HJD & RV & RV$_{err}$  \\
    & km~s$^{-1}$ & km~s$^{-1}$\\
\hline
2454793.55000 &  $+$346.064 & 0.053\\
2454820.45944 &  $+$349.034 & 0.025\\
2455126.69729 &  $+$342.750 & 0.015\\
2455171.57052 &  $+$343.843 & 0.014\\
2455207.51291 &  $+$347.058 & 0.018\\
2455439.72402 &  $+$349.472 & 0.014\\
2455478.72231 &  $+$345.614 & 0.052\\
2455503.57008 &  $+$343.789 & 0.046\\
2455531.52335 &  $+$342.751 & 0.030\\
2455860.69191 &  $+$347.747 & 0.015\\
2455892.55739 &  $+$344.930 & 0.117\\
2456191.75734 &  $+$354.904 & 0.012\\
2456209.72120 &  $+$353.449 & 0.021\\
2456531.71198 &  $+$356.770 & 0.018\\
2456917.65964 &  $+$356.020 & 0.025\\
\hline                           
\end{tabular}           
\end{table}            

\begin{table}                   
\caption{HE~0854$+$0151}         
\centering                       
\begin{tabular}{lcc}
\hline\hline
HJD & RV & RV$_{err}$  \\
    & km~s$^{-1}$ & km~s$^{-1}$\\
\hline
2454516.50743 &  $+$135.575 & 0.033\\
2454780.70470 &  $+$144.261 & 0.072\\
2454930.38332 &  $+$130.585 & 0.034\\
2454964.39430 &  $+$124.541 & 0.054\\
2455149.73265 &  $+$141.161 & 0.032\\
2455171.71750 &  $+$144.308 & 0.024\\
2455174.67146 &  $+$144.662 & 0.041\\
2455232.52829 &  $+$145.394 & 0.025\\
2455531.69023 &  $+$139.809 & 0.040\\
2455620.45671 &  $+$145.626 & 0.041\\
2455662.42371 &  $+$140.074 & 0.028\\
2455822.74479 &  $+$121.901 & 0.112\\
2455915.75991 &  $+$138.642 & 0.129\\
2456005.48541 &  $+$145.867 & 0.061\\
2456273.71422 &  $+$132.054 & 0.062\\
\hline                           
\end{tabular}           
\end{table}            

\clearpage

\begin{table}                   
\caption{HE~0959$-$1424}         
\centering                       
\begin{tabular}{lcc}
\hline\hline
HJD & RV & RV$_{err}$  \\
    & km~s$^{-1}$ & km~s$^{-1}$\\
\hline
2454406.77276   &  $+$344.076 & 0.171\\
2454793.71150   &  $+$344.176 & 0.409\\
2455149.76314   &  $+$343.803 & 0.156\\
2455174.72219   &  $+$343.757 & 0.181\\
2455207.58849   &  $+$343.775 & 0.145\\
2455310.37042   &  $+$343.598 & 0.137\\
2455544.72364   &  $+$343.630 & 0.204\\
2455620.54421   &  $+$343.708 & 0.140\\
2455662.38833   &  $+$343.659 & 0.128\\
2455944.56413   &  $+$343.632 & 0.246\\
2456005.51849   &  $+$343.377 & 0.283\\
2456033.36757   &  $+$343.459 & 0.357\\
2456399.38501   &  $+$343.025 & 0.248\\
2456722.52147   &  $+$342.309 & 0.519\\
2456756.38762   &  $+$342.602 & 0.268\\
2456796.37413   &  $+$342.613 & 0.174\\
2457142.38500   &  $+$341.818 & 0.273\\
\hline
\end{tabular}           
\end{table}            

\begin{table}                   
\caption{HE~1045$+$0226}         
\centering                       
\begin{tabular}{lcc}
\hline\hline
HJD & RV & RV$_{err}$  \\
    & km~s$^{-1}$ & km~s$^{-1}$\\
\hline
2456307.69698 &  $+$131.672 & 0.228\\
2456399.43421 &  $+$131.652 & 0.099\\
2456652.61751 &  $+$131.564 & 0.104\\
2456712.67531 &  $+$131.685 & 0.364\\
2457076.61932 &  $+$131.460 & 0.399\\
2457110.44369 &  $+$130.954 & 0.178\\
\hline                           
\end{tabular}                    
\end{table}            

\begin{table}                   
\caption{HE~1046$-$1352}         
\centering                       
\begin{tabular}{lcc}
\hline\hline
HJD & RV & RV$_{err}$  \\
    & km~s$^{-1}$ & km~s$^{-1}$\\
\hline
2454909.61389 &  $+$94.395 & 0.795\\
2454930.45563 &  $+$97.658 & 0.147\\
2454964.43469 &  $+$47.693 & 1.670\\
2455174.74220 &  $+$105.913 & 0.851\\
2455232.57334 &  $+$96.330 & 0.147\\
2455310.39271 &  $+$73.223 & 0.859\\
2455344.39166 &  $+$48.887 & 0.278\\
2455554.78013 &  $+$94.567 & 0.543\\
2455620.51605 &  $+$97.700 & 0.553\\
2456006.50365 &  $+$71.838 & 0.545\\
2456033.41230 &  $+$51.174 & 1.102\\
2456721.61189 &  $+$74.268 & 1.555\\
\hline                   
\end{tabular}            
\end{table}              
               
\begin{table}                    
\caption{CS~30301$-$015}            
\centering                       
\begin{tabular}{lcc}
\hline\hline
HJD & RV & RV$_{err}$  \\
    & km~s$^{-1}$ & km~s$^{-1}$\\
\hline
2454254.62579  &  $+$86.747 & 0.094\\
2454314.47889  &  $+$86.510 & 0.081\\
2454480.79001  &  $+$86.638 & 0.054\\
2454625.52411  &  $+$86.607 & 0.049\\
2454909.59318  &  $+$86.585 & 0.078\\
2454930.71839  &  $+$86.636 & 0.066\\
2454951.73429  &  $+$86.674 & 0.094\\
2454987.49075  &  $+$86.577 & 0.048\\
2455059.37482  &  $+$86.462 & 0.055\\
2455344.59436  &  $+$86.562 & 0.048\\
2455415.40116  &  $+$86.663 & 0.058\\
2455439.37958  &  $+$86.510 & 0.056\\
2455704.56070  &  $+$86.620 & 0.065\\
2455776.46938  &  $+$86.611 & 0.064\\
2456005.71388  &  $+$86.564 & 0.134\\
2456033.63882  &  $+$86.633 & 0.058\\
2456078.51945  &  $+$86.566 & 0.061\\
2456488.39989  &  $+$86.763 & 0.044\\
\hline
\end{tabular}           
\end{table}            

\begin{table}                   
\caption{HE~1523$-$1155}         
\centering                       
\begin{tabular}{lcc}
\hline\hline
HJD & RV & RV$_{err}$  \\
    & km~s$^{-1}$ & km~s$^{-1}$\\
\hline
2456756.69769   &  $-$48.312 & 0.034\\
2456796.67508   &  $-$46.377 & 0.051\\
2456888.38390   &  $-$40.622 & 0.040\\
2457110.72622   &  $-$46.063 & 0.069\\
2457142.65599   &  $-$44.028 & 0.037\\
2457168.61410   &  $-$42.363 & 0.026\\
2457225.44781   &  $-$39.130 & 0.042\\
2457239.39307   &  $-$38.552 & 0.020\\
2457258.36919   &  $-$38.016 & 0.042\\
\hline                           
\end{tabular}                    
\end{table}            

\begin{table}                   
\caption{HE~2201$-$0345}         
\centering                       
\begin{tabular}{lcc}
\hline\hline
HJD & RV & RV$_{err}$  \\
    & km~s$^{-1}$ & km~s$^{-1}$\\
\hline
2454314.59436   &  $-$54.804 & 0.053\\
2454338.51241   &  $-$54.763 & 0.098\\
2454373.44648   &  $-$54.627 & 0.078\\
2454665.70652   &  $-$53.021 & 0.093\\
2454964.70705   &  $-$50.946 & 0.142\\
2455009.67131   &  $-$50.441 & 0.093\\
2455059.48210   &  $-$50.167 & 0.071\\
2455126.49880   &  $-$49.927 & 0.067\\
2455344.66126   &  $-$51.510 & 0.053\\
2455415.47351   &  $-$52.802 & 0.062\\
2455439.41067   &  $-$53.130 & 0.051\\
2455503.33221   &  $-$54.321 & 0.155\\
2455531.36578   &  $-$54.418 & 0.185\\
2455704.66231   &  $-$56.625 & 0.042\\
2455776.49350   &  $-$57.152 & 0.086\\
2455882.30261   &  $-$58.114 & 0.092\\
2455898.30389   &  $-$58.260 & 0.091\\
2456033.73856   &  $-$58.053 & 0.236\\
2456078.70413   &  $-$58.791 & 0.096\\
2456139.62139   &  $-$58.890 & 0.102\\
2456518.50427   &  $-$59.730 & 0.532\\
2456530.61456   &  $-$59.760 & 0.062\\
2456603.44706   &  $-$59.703 & 0.121\\
2456886.51808   &  $-$59.808 & 0.085\\
2456956.46129   &  $-$60.095 & 0.197\\
2457192.70683   &  $-$60.170 & 0.059\\
2457257.54595   &  $-$60.008 & 0.100\\
\hline
\end{tabular}           
\end{table}

\begin{table}                   
\caption{HE~2312$-$0758}         
\centering                       
\begin{tabular}{lcc}
\hline\hline
HJD & RV & RV$_{err}$  \\
    & km~s$^{-1}$ & km~s$^{-1}$\\
\hline
2456191.43803    &  $+$38.553 & 0.050\\
2456213.45637    &  $+$38.391 & 0.054\\
2456488.71018    &  $+$34.121 & 0.045\\
2456518.67707    &  $+$33.749 & 0.051\\
2456528.66775    &  $+$33.640 & 0.063\\
2456545.61190    &  $+$33.228 & 0.046\\
2456887.52097    &  $+$30.174 & 0.067\\
2456917.53578    &  $+$30.435 & 0.073\\
2456956.48012    &  $+$30.255 & 0.074\\
2457225.62292    &  $+$30.102 & 0.062\\
2457257.56346    &  $+$30.139 & 0.048\\
\hline                           
\end{tabular}                    
\end{table}                      
                                 
\begin{table}                    
\caption{HE~2330$-$0555}            
\centering                       
\begin{tabular}{lcc}
\hline\hline
HJD & RV & RV$_{err}$  \\
    & km~s$^{-1}$ & km~s$^{-1}$\\
\hline
2454314.711836  &  $-$234.968 & 0.046\\
2454373.550888  &  $-$235.089 & 0.052\\
2455059.605701  &  $-$235.218 & 0.054\\
2455080.608758  &  $-$235.039 & 0.108\\
2455126.539073  &  $-$235.182 & 0.059\\
2455176.414354  &  $-$235.118 & 0.142\\
2455415.633807  &  $-$235.179 & 0.052\\
2455439.497335  &  $-$235.104 & 0.060\\
2455478.573649  &  $-$235.114 & 0.078\\
2455531.439197  &  $-$235.126 & 0.094\\
2455776.707594  &  $-$235.231 & 0.070\\
2455796.607358  &  $-$235.370 & 0.091\\
2455882.421279  &  $-$235.048 & 0.045\\
2456141.696801  &  $-$234.927 & 0.092\\
2456241.363645  &  $-$234.912 & 0.212\\
2456530.642253  &  $-$235.139 & 0.066\\
2456887.546840  &  $-$235.386 & 0.094\\
\hline          
\end{tabular}           
\end{table}

\begin{table}                   
\caption{HE~0039$-$2635}         
\centering                       
\begin{tabular}{lcc}
\hline\hline
HJD & RV & RV$_{err}$  \\
    & km~s$^{-1}$ & km~s$^{-1}$\\
\hline
2456241.41669 & $-$52.078 & 0.023\\
2456518.71254 & $-$43.400 & 0.023\\
\hline
\end{tabular}           
\end{table}

\begin{table}                    
\caption{HE~1031$-$0020}            
\centering                       
\begin{tabular}{lcc}
\hline\hline
HJD & RV & RV$_{err}$  \\
    & km~s$^{-1}$ & km~s$^{-1}$\\
\hline
2454219.42938  &  $+$69.718 & 0.103\\
2454254.44108  &  $+$69.319 & 0.152\\
2454406.74096  &  $+$69.519 & 0.119\\
2454459.64146  &  $+$69.839 & 0.141\\
2454464.63446  &  $+$69.944 & 0.086\\
2454481.61235  &  $+$69.524 & 0.145\\
2454516.55332  &  $+$69.634 & 0.096\\
2454793.76871  &  $+$69.565 & 0.136\\
2454909.51656  &  $+$69.473 & 0.124\\
2454930.53075  &  $+$69.308 & 0.067\\
2455174.70264  &  $+$69.198 & 0.095\\
2455207.60630  &  $+$69.142 & 0.075\\
2455544.74047  &  $+$68.903 & 0.114\\
2455620.49236  &  $+$68.592 & 0.096\\
2455915.79512  &  $+$68.460 & 0.222\\
2455944.58281  &  $+$67.917 & 0.197\\
2456033.38640  &  $+$68.077 & 0.098\\
2456241.74480  &  $+$67.609 & 0.138\\
2456399.40659  &  $+$67.486 & 0.120\\
2456722.48940  &  $+$66.819 & 0.149\\
2457110.39343  &  $+$66.361 & 0.227\\
2457142.44270  &  $+$66.111 & 0.106\\
\hline             
\end{tabular}           
\end{table}            
                   
\begin{table}                    
\caption{LP~625$-$44}            
\centering                       
\begin{tabular}{lcc}
\hline\hline
HJD & RV & RV$_{err}$  \\
    & km~s$^{-1}$ & km~s$^{-1}$\\
\hline
2454219.72236  &  $+$38.955 & 0.039\\
2454254.65878  &  $+$38.870 & 0.111\\
2454285.56286  &  $+$39.027 & 0.060\\
2454314.49424  &  $+$38.911 & 0.041\\
2454338.41720  &  $+$38.770 & 0.047\\
2454373.35738  &  $+$38.773 & 0.051\\
2454625.62936  &  $+$38.051 & 0.038\\
2454665.57095  &  $+$38.081 & 0.093\\
2454930.74622  &  $+$37.306 & 0.046\\
2454964.69845  &  $+$37.168 & 0.059\\
2454987.52690  &  $+$37.153 & 0.036\\
2455059.43020  &  $+$36.853 & 0.075\\
2455344.60471  &  $+$36.007 & 0.060\\
2455415.37378  &  $+$35.652 & 0.057\\
2455439.39220  &  $+$35.608 & 0.057\\
2455620.75079  &  $+$34.869 & 0.059\\
2455704.58037  &  $+$34.615 & 0.062\\
2455776.41360  &  $+$34.301 & 0.048\\
2456005.74059  &  $+$33.446 & 0.086\\
2456033.68185  &  $+$33.154 & 0.106\\
2456062.71485  &  $+$33.086 & 0.091\\
2456078.55749  &  $+$32.913 & 0.069\\
2456488.41291  &  $+$30.863 & 0.052\\
2456574.33217  &  $+$30.479 & 0.037\\
2456579.32582  &  $+$30.406 & 0.081\\
2456756.74336  &  $+$29.588 & 0.167\\
2456796.71349  &  $+$29.225 & 0.068\\
2456887.37991  &  $+$28.872 & 0.056\\
\hline                   
\end{tabular}            
\end{table}              
                         
\clearpage
                         
\section{Literature data for the single programme stars}

\begin{table}[ht]
\caption{Mean heliocentric radial velocities from the literature and total
  time-span covered for the single stars }
\label{tbl-B1}
\centering
\begin{tabular}{lrrrrl}
\hline\hline
Stellar ID & $\Delta$T Total &  mean RV (this work)  & mean RV & N & Ref \\
           & (days)          &(km~s$^{-1}$)          & (km~s$^{-1}$)&   &  \\
\hline
\object{HE~0206$-$1916}& 4445 &$-$199.509 &$-$200.0 &1 &\citet{aoki2007} \\
\hline
\object{HE~1045$+$0226}& 1868 &$+$131.498 &$+$131.2 &1 &\citet{cohen2013}\\ 
\hline
\object{HE~2330$-$0555}& 4314 &$-$235.124 &$-$235.0 &1 & \citet{aoki2007}\\
\hline
\object{CS~30301$-$015}& 4352 &$+$86.61  &$+$85.5 &1& \citet{tsangarides2003}   \\
                       &      &          &$+$86.5 &1& \citet{aoki2002c} \\
                       &      &          &$+$85.5 &2& \citet{lucatello2005}\\
\hline
\end{tabular}
\end{table}

\end{appendix}           
                         

\begin{thebibliography}{}

\bibitem[Abate et al.(2013)]{abate2013} Abate, C., Pols, O. R., Izzard, R. G.,
  Mohamed, S. S., \& de Mink, S. E., 2013, \aap, 552, A26

\bibitem[Abate et al.(2015a)]{abate2015a} Abate, C., Pols, O. R., Karakas, A.,
  Izzard, R. G., 2015a, \aap, 576, A118

\bibitem[Abate et al.(2015b)]{abate2015b} Abate, C., Pols, O. R., 
  Stancliffe, R. J., et al. 2015b, \aap, 581, 62

\bibitem[Allen et al.(2012)]{allen2012} Allen, D. M., Ryan, S. G., 
  Rossi,S., Beers, T. C., \& Tsangarides, S. A. 2012, \aap, 548, 34

\bibitem[Aoki et al.(2000)]{aoki2000} Aoki, W., Norris, J. E., Ryan, S. G.,
  Beers, T. C., \& Ando, H., 2000, \apjl, L97

\bibitem[Aoki et al.(2002a)]{aoki2002a} Aoki, W., Ryan, S. G., Beers, T. C., \&
  Ando, H., 2002a, \pasj, 54, 933

\bibitem[Aoki et al.(2002b)]{aoki2002b} Aoki, W., Ryan, S. G., Norris, J. E.,
  Beers, T. C., Ando, H., \& Tsangarides, S. A., 2002b, \apj, 580, 1149

\bibitem[Aoki et al.(2002c)]{aoki2002c} Aoki, W., Ando, H., Honda, S., et al.,
  2002c, \pasj, 54, 427

\bibitem[Aoki et al.(2007)]{aoki2007} Aoki, W., Beers, T. C., Norris, J. E.,
  Ryan, S. G., \& Tsangarides, S. A., 2007, \apj, 655, 492

\bibitem[Barbuy et al.(2005)]{barbuy2005} Barbuy, B., Spite, M., Spite, F., et
  al., 2005, \aap, 429, 1031

\bibitem[Beers et al.(1985)]{beers1985} Beers, T. C., Preston, G. W.,\&
  Shectman, S. A., 1985, \aj, 90, 2089

\bibitem[Beers et al.(1992)]{beers1992} Beers, T. C., Preston, G. W.,\&
  Shectman, S. A., 1992, \aj, 103, 1987

\bibitem[Beers \& Christlieb(2005)]{beerschristlieb2005} Beers, T. C., \&
  Christlieb, N., 2005, \araa, 43, 531

\bibitem[Beers et al.(2007a)]{beers2007a} Beers, T. C., Flynn, C., Rossi, S., et
  al., 2007a, \apjs, 168, 128

\bibitem[Beers et al.(2007b)]{beers2007b} Beers, T. C., Sivarani, T., Marsteller,
  B., Lee, Y., Rossi, S., \& PLez, B., 2007b, \aj, 133, 1193

\bibitem[Beers et al.(2014)]{beers2014} Beers, T. C., Norris, J. E., Placco, V. M.,
et al., 2014, \apj, 794, 58

\bibitem[Bertolli et al.(2013)]{bertolli2013} Bertolli, M. G., Herwig, F., 
  Pignatari, M., \& Kawano, T., 2013, (arXiv:1310.4578)

\bibitem[Bisterzo et al.(2012)]{bisterzo2012} Bisterzo, S., Gallino, R.,
  Straniero, O., Cristallo, S., \& K{\"a}ppeler, F., 2012, \mnras, 422, 849 

\bibitem[Boffin \& Jorissen(1988)]{boffin1988} Boffin, H. M. J., \& Jorissen,
  A., 1988, \aap, 205, 155

\bibitem[Bondi \& Hoyle(1944)]{bondihoyle1944} Bondi, H., \& Hoyle, F., 1944,
  \mnras, 104, 273

\bibitem[Bonifacio et al.(2015)]{bonifacio2015} Bonifacio, P., Caffau, E.,
  Spite, M., et al., 2015, \aap, 579, A28

\bibitem[Carney et al. (2003)]{carney2003} Carney, B. W., Latham, D. W.,
  Stefanik, R. P., et al., 2003, \aj, 125, 293

\bibitem[Christlieb et al.(2008)]{christlieb2008} Christlieb, N., Sch{\"o}rck,
  T., Frebel, A., et al., 2008, \aap, 484, 721

\bibitem[Cohen et al.(2003)]{cohen2003} Cohen, J. G., Christlieb, N., Qian, Y,
  Z., \& Wasserburg, G. J., 2003, \apj, 588, 1082

\bibitem[Cohen et al.(2006)]{cohen2006} Cohen, J. G., McWilliam, A., Shectman, 
  S., et al., 2006, \aj, 132, 137

\bibitem[Cohen et al.(2013)]{cohen2013} Cohen, J. G., Christlieb, N., Thompson,
  I., et al., 2013, \apj, 778, 56

\bibitem[Cowan \& Rose (1977)]{cowan1977} Cowan, J. J., \& Rose, W. K., 1977,
  \apj, 212, 149

\bibitem[Cristallo et al.(2011)]{cristallo2011} Cristallo, S., Piersanti, L.,
  Straniero, O., et al., 2011, \apjs, 197, 17 

\bibitem[Cristallo et al.(2009)]{cristallo2009} Cristallo, S., Straniero, O.,
  Gallino, R., et al., 2009, \apj, 696, 797

\bibitem[Duquennoy \& Mayor (1991)]{duquennoy91} Duquennoy, A., \& 
  Mayor, M., 1991, \aap, 248, 485

\bibitem[Frebel \& Norris (2015)]{frebelnorris2015} Frebel, A., \& Norris,
  J. E., 2015, \araa, 53, 631 

\bibitem[Frischknecht et al.(2012)]{frischknecht2012} Frischknecht, U., Hirschi, 
  R., \& Thielemann, F.-K., 2012, \aap, 538, L2

\bibitem[Frischknecht et al.(2015)]{frischknecht2015} Frischknecht, U., Hirschi, 
  R., Pignatari, M., et al., 2015, \mnras, in press (arXiv 1511.05730)

\bibitem[Goriely \& Siess (2005)]{goriely2005} Goriely, S., \& Siess, L.,
  2005, IAU Symposium 228, From Lithium to Uranium: Elemental Tracers of Early
  Cosmic Evolution, ed. Hill, V., Francois, P., \& Primas, F., 451

\bibitem[Hansen et al.(2011)]{hansen2011} Hansen, T., Andersen, J., Nordstr{\"o}m, 
  B., Buchhave, L., \& Beers, T. C., 2011, \apj, 743, L1 

\bibitem[Hansen et al.(2015a)]{hansen2015a} Hansen, T. T., Hansen, C. J.,
  Christlieb, N., et al., 2015a, \apj, 807, 173

\bibitem[Hansen et al.(2015b)]{hansen2015b} Hansen, T. T., Andersen, J.,
  Nordstr{\"o}m, B., et al., 2015b, \aap, 583, 49 (Paper I)

\bibitem[Hansen et al.(2015c)]{hansen2015c} Hansen, T. T., Andersen, J.,
  Nordstr{\"o}m, B., et al., 2015c, \aap, in press (arXiv 1511.08197) (Paper II)

\bibitem[Henden et al.(2015)]{henden2015} Henden, A. A., Levine, S., Terrell,
  D., \& Welch, D. L., 2015, American Astronomical Society Meeting Abstracts, 225, 336.16

\bibitem[Hirschi(2007)]{hirschi2007} Hirschi, R., 2007, \aap, 461, 571

\bibitem[Ivezic et al.(2012)]{ivezic2012} Ivezic, Z., Beers, T. C., \& Juric, M.
2012, \araa, 50, 251

\bibitem[Jorissen et al.(1998)]{jorissen1998} Jorissen, A., Van Eck, S.,
  Mayor, M., \& Udry, S., 1988, \aap, 332, 877 

\bibitem[Jorissen et al.(2015a)]{jorissen2015a} Jorissen, A., Hansen, T., Van 
  Eck, S., et al., 2015a, \aap, in press (arXiv 1510.06045)

\bibitem[Jorissen et al.(2015b)]{jorissen2015b} Jorissen, A., Van Eck, S.,
  Van Winkel, H., et al., 2015b, \aap, in press (arXiv 1510.05840) 

\bibitem[Karakas(2010)]{karakas2010} Karakas, A.~I.\ 2010, \mnras, 403, 1413 

\bibitem[Karakas \& Lattanzio (2014)]{karakas2014} Karakas, A. I, \&
  Lattanzio, J. C., 2014, \pasa, 31, 30

\bibitem[Kennedy et al.(2011)]{kennedy2011} Kennedy, C. R., Sivarani, T.,
  Beers., T. C., 2011, \aj, 141, 102

\bibitem[Kopal (1959)]{kopal1959} Kopal, Z., 1959, Close Binary Systems, The
  International Astrophysics Series, London: Chapman Hall

\bibitem[Lucatello et al.(2003)]{lucatello2003} Lucatello, S, Gratton, R.,
  Cohen, J. G., et al., 2003, \aj, 125, 875

\bibitem[Lucatello et al.(2005)]{lucatello2005} Lucatello, S., Tsangarides,
  S. A., Beers, T. C., Carretta, E., Gratton, R. G., \& Ryan, S. G., 2005, \apj,
  625, 825

\bibitem[Lugaro et al.(2012)]{lugaro2012} Lugaro, M., Karakas, A. I.,
  Stancliffe, R. J., \& Rijs, C., 2012, \apj, 747, 2 

\bibitem[Maeder et al.(2015)]{maeder2015} Maeder, A., Meynet, G., \&
  Chiappini, C., 2015, \aap, 576, A56

\bibitem[Masseron et al.(2010)]{masseron2010} Masseron, T., Johnson, J. A.,
  Plez, B., et al., 2010, \aap, 509, A93

\bibitem[Mathieu et al.(1990)]{mathieu1990} Mathieu, R. D., Latham, D. W., 
  \& Griffin, R. F., 1990, \aj, 100, 1899

\bibitem[McClure \& Woodsworth (1990)]{mcclure1990} McClure, R. D., \& 
  Woodsworth, A. W., 1990, \apj, 352, 709

\bibitem[Merle et al.(2015)]{merle2015} Merle, T., Jorissen, A., Van Eck, S.,
  Maseron, T., \& Van Winkel, H., 2015, \aap, subm. (arXiv 1510.05908)

\bibitem[Mermilliod et al.(2007)]{mermio2007} Mermilliod, J.-C., Andersen, J.,
  Latham, D. W., \& Mayor, M., 2007, \aap, 473, 829 

\bibitem[Meynet et al.(2006)]{meynet2006} Meynet, G., Ekstr{\"o}m,  S., 
  \& Maeder, A., 2006, \aap, 447, 623

\bibitem[Mohamed \& Podsiadlowsky (2007)]{mohamed2007} Mohamed, S., \&
  Podsiadlowski, P., 2007, Astronomical Society of the Pacific Conference
  Series, 372, ed. Napiwotzki, R., \& Burleigh, M. R., 397

\bibitem[Morbey \& Griffin(1987)]{morbeygriffin1987} Morbey, C. L. \& 
  Griffin, R. F., 1987, \apj, 317, 343

\bibitem[Nordstr{\"o}m et al.(1997)]{nordstrom1997} Nordstr{\"o}m, B., Stefanik, 
   R. P., Latham, D. W., Andersen, J., 1997, \aaps, 126, 21 

\bibitem[Norris et al.(1997)]{norris1997} Norris, J. E., Ryan, S. G., \&
  Beers, T. C., 1997, \apj, 488, 350

\bibitem[Paczy{\'n}ski (1965)]{paczynski1965} Paczy{\'n}ski, B., 1965, \actaa,
  15, 89

\bibitem[Paczy{\'n}ski (1976)]{paczynski1976} Paczy{\'n}ski, B., 1976, IAU Symposium 73,
  Structure and Evolution of Close Binary Systems, ed. Eggleton, P., Mitton,
  S., \& Whelan, J., 75

\bibitem[Pignatari et al.(2008)]{pignatari2008} Pignatari, M., Gallino, R., 
  Meynet, G., et al., 2008, \apj, 687, L95

\bibitem[Placco et al.(2010)]{placco2010} Placco, V. M., Kennedy, C. R., Rossi, S.,
et al., 2010, \aj, 139, 1051

\bibitem[Placco et al.(2011)]{placco2011} Placco, V. M., Kennedy, C. R., Beers,
  T. C., et al., 2011, \aj, 142, 188

\bibitem[Placco et al.(2013)]{placco2013} Placco, V.~M., Frebel, 
  A., Beers, T.~C., et al., 2013, \apj, 770, 104

\bibitem[Placco et al.(2015)]{placco2015} Placco, V.~M., Beers, 
  T.~C., Ivans, I.~I., et al., 2015, \apj, in press (arXiv:1508.05872) 

\bibitem[Preston \& Sneden (2000)]{preston2000} Preston, G. W., \& Sneden, C.,
  2000, \aj, 120, 1014

\bibitem[Preston \& Sneden (2001)]{preston2001} Preston, G. W., \& Sneden, C.,
  2001, \aj, 122, 1545

\bibitem[Richard et al. (2002)]{richard2002} Richard, O., Michaud, G., Richer,
  J., Turcotte, S., Turck-Chi{\`e}ze, S., \& Vandenberg, D. A., 2002, \apj,
  568, 979

\bibitem[Ricker \& Taam (2008)]{ricker2008} Ricker, P. M., \& Taam, R. E.,
  2008, \apjl, 672, L41

\bibitem[Riebel et al.(2010)]{riebel2010} Riebel, D., Meixner, M., Fraser, O. 
   et al., \apj, 723, 1195

\bibitem[Ryan et al.(2005)]{ryan2005} Ryan, S. G., Aoki, W., Norris, J. E., \&
  Beers, T. C., 2005, \apj, 635, 349

\bibitem[Spite et al.(2013)]{spite2013} Spite, M., Caffau, E., Bonifacio, P.,
  et al., 2013, \aap, 552, A107 

\bibitem[Stancliffe \& Glebbeek(2008)]{stancliffe2008} Stancliffe, R., \&
  Glebbeek, E., 2008, \mnras, 389, 1828

\bibitem[Stancliffe et al.(2007)]{stancliffe2007} Stancliffe, R., Glebbeek,
  E., Izzard, R. G., \& Pols., O.R., 2007, \aap, 464, L57

\bibitem[Starkenburg et al.(2014)]{starkenburg2014} Starkenburg, E., Shetrone,
  M. D., McConnachie, A. W., \& Venn, K. A., 2015, \mnras, 441, 1217

\bibitem[Tsangarides et al.(2003)]{tsangarides2003} Tsangarides, S. A., Ryan,
  S. G., \& Beers, T. C., 2003, Astronomical Society of the Pacific Conference
  Series, ed: Charbonnel, C., Schaerer, D., \& Meynet, G., 304, 133

\bibitem[Umeda \& Nomoto(2003)]{umeda2003} Umeda, H., \& Nomoto, K., 2003,
  \nat, 422, 871

\bibitem[Vassiliadis \& Wood(1993)]{vassiliadis1993} Vassiliadis, E. \& Wood, P. R. 1993, \apj, 413, 641

\bibitem[Yong et al.(2013)]{yong2013} Yong, D., Norris, J.E., Bessell, M. S.,
  et al., 2013, \apj, 762, 26


\end{thebibliography}
\end{document}